\documentclass{ws-book9x6}
\usepackage{ws-book-thm}   
\usepackage{ws-book-van}
\usepackage[pdfpagelabels=false,colorlinks=true,allcolors=black]{hyperref}  
\usepackage{sidecap}

\usepackage[T1, T2A]{fontenc}
\usepackage{hyperref}

\title{Spin Glass Theory and Further Beyond}      

\renewcommand{\thechapter}{3}

\makeindex

\begin{document}
\chapter{Beyond  Ising SK} 

\label{ch3A}
\vspace{-2.3cm}
\begin{center}
\small
{\bf{$m$-vector, Potts, $p$-spin, spherical, induced moment, random graphs}}
\normalsize
\vspace{0.5 cm}
\end{center}
\author{ David Sherrington}   
\address {Rudolf Peierls Centre for Theoretical Physics, Oxford OX1 3PU, UK}
\author{Jairo R. L. de Almeida}
\address{Departamento de Física, Universidade Federal de Pernambuco, Recife, Brazil}
\vspace{0.5 cm}


This chapter introduces a variety of spin glass models, beyond the simple Sherrington-Kirkpatrick (SK) version, that have led to  enriched understanding and application. 


\section{Introduction}

When one of us introduced the SK model \cite{SK75}, with Kirkpatrick in 1975,  following the very innovative paper of Edwards and Anderson
\cite{EA1975} earlier the same year, 
he made two choices that have proven to be influential;
 (i) an infinite-range intensive exchange distribution and (ii)  Ising spins; the first in the hope of enabling an exact solution by  analogy with and extension of previously known work on pure magnets
and to provide the correct mean-field solution; the second to simplify the technicalities. In the event, the correct mean-field solution turned out to be much more subtle and interesting than was anticipated and thus revealing that mean-field theory need not be trivial -- a common perspective at the time. The use of discrete Ising spins was serendipitous in highlighting a problem with a `natural' (replica-symmetric) ansatz, recognized  through an
unphysical negative entropy feature already in SK. (The problem would not have been obvious for continuous classical spins where such a feature is a known pathology\footnote{EA  \cite{EA1975} considered $m=3$ spins, presumably because the stimulating experiments at the time involved Heisenberg spins.}.) The probable origin of the unphysical behaviour was later shown by the demonstration of an instability in an excitation mode in replica-space by the other of us and Thouless \cite{Almeida1978a}, eventually  leading Parisi between 1979 \cite{Parisi1979} and 1983 \cite{Parisi1983} to the correct and highly impactful (full) replica symmetry breaking (RSB) resolution.

In this chapter we introduce systems with spins and interactions of different symmetries and indicate some of the qualitatively new consequences that arise. Mainly, we concentrate on 
systems with quenched infinite-range interactions, drawn randomly and independently from identical distributions, in conceptual analogy with SK, in the expectation of solvability in the thermodynamic limit $N \rightarrow \infty$ and the definition of the appropriate associated mean-field theories.
We concentrate on equilibrium/Gibbsian statistical mechanics (statics), leaving dynamics to later chapters.

\section{Binary interaction models}

\subsection{Local moment spins of $O(m)$ symmetry}

In this section we consider systems with classical $m$-vector spins ${\bf{S}}_i$ ($m$ integral) beyond Ising ($m=1$) and with 
SK-like scalar-product interactions. 

In the absence of anisotropy these have the Hamiltonian
\begin{equation}
H= -\sum_{(ij)}  J_{ij} {\bf{S}}_{i}.{\bf{S}}_{j},
\label{SK}
\end{equation}
with the $J_{ij} $  drawn independently randomly and symmetrically ($J_{ij}=J_{ji}$) {\it{\`{a} la} }SK, from a Gaussian distribution of mean\footnote{The inclusion of a finite positive mean, originally by Sherrington and Southern \cite{Sherrington-Southern}, aims  to emulate the concentration-dependence of experimental spin glass systems.  } $J_{0}/N$ and variance  $J^{2}/{N}$.  
The subsequent behaviour is essentially as for the $m=1$ case (For RS analysis see \cite{Kirkpatrick1978} and \cite{Almeida1978}).  In particular, if the spin length normalisation is chosen as 
$|{\bf{S}}| =\sqrt{m}$ the spin glass transition temperature (in units with Boltzmann's constant set to unity, $k_{B}=1$) is $T=J$.

Something new arises when anisotropic terms are included.

\begin{figure}[ht]
\centering
\sidebyside
{\includegraphics[width=1.9in]{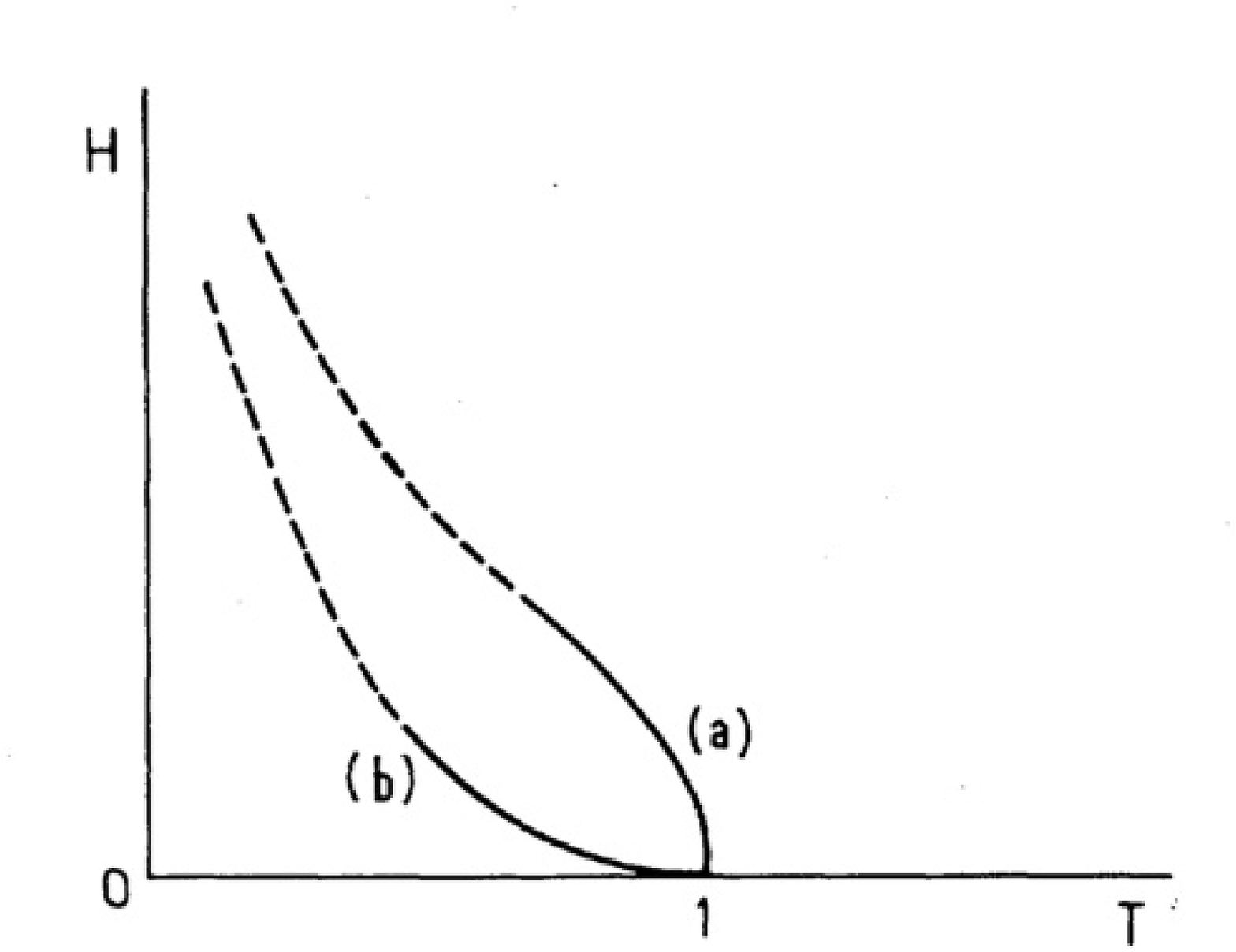}\\(i)} 
{\includegraphics[width=2.1in]{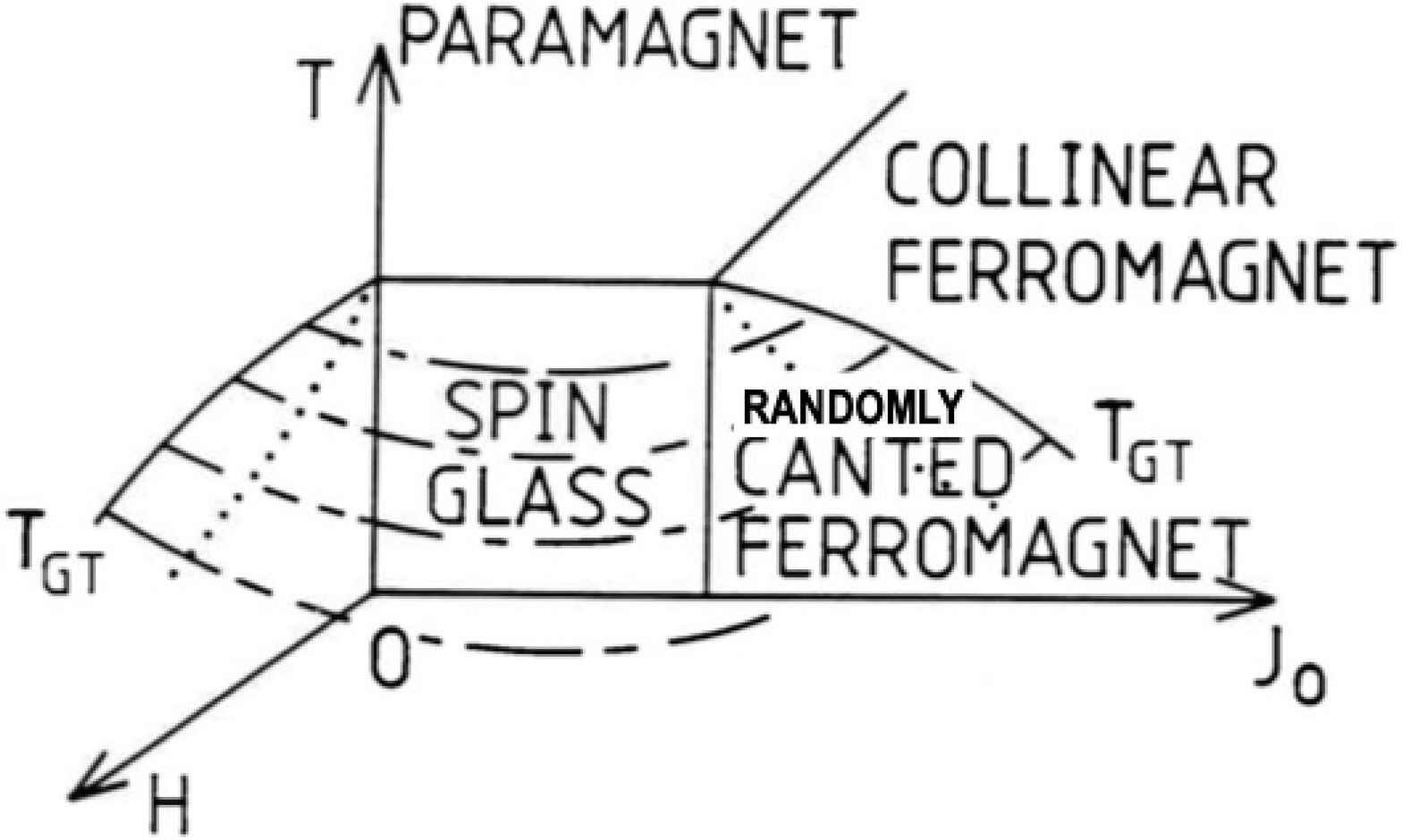}\\(ii)}
\caption{$m$-vector phase diagrams: (i) $J_0=0$ in an applied field H: (a) GT transverse spin-glass line, (b) AT crossover (figure from \cite{Gabay1981},with permission. \copyright  American Physical Society(1981),
 (ii)  with also a positive  exchange distribution mean $J_0$; solid lines denote phase transitions, dashed lines denote quasi-AT  crossovers (figure adapted from \cite{Sherrington1983}, with permission \copyright Physical Society of Japan (1983)).}.
 \label{GT}

\sidebyside
{\includegraphics[width=1.9 in]{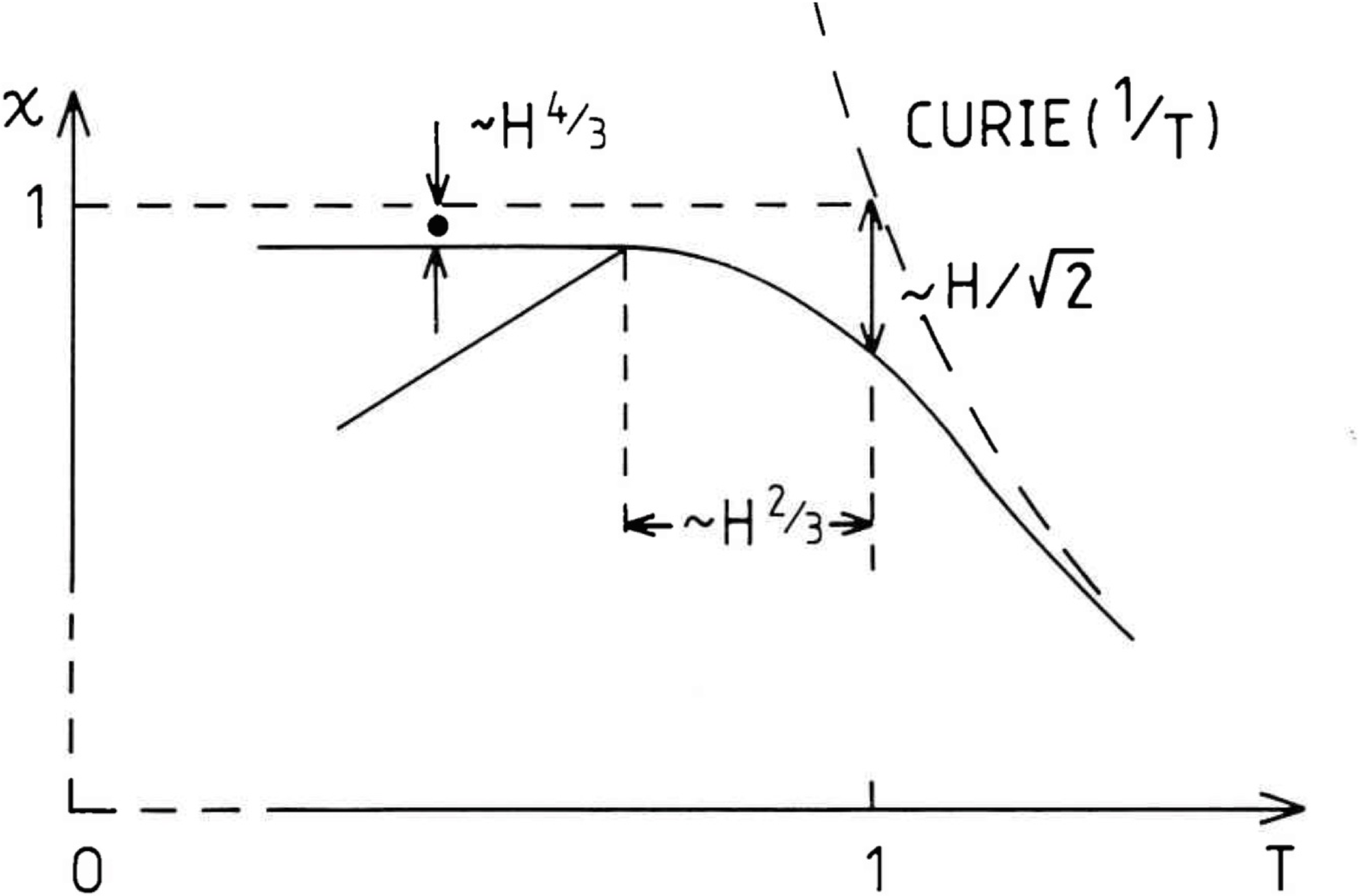}\\(a)}
{\includegraphics[width=2.1 in]{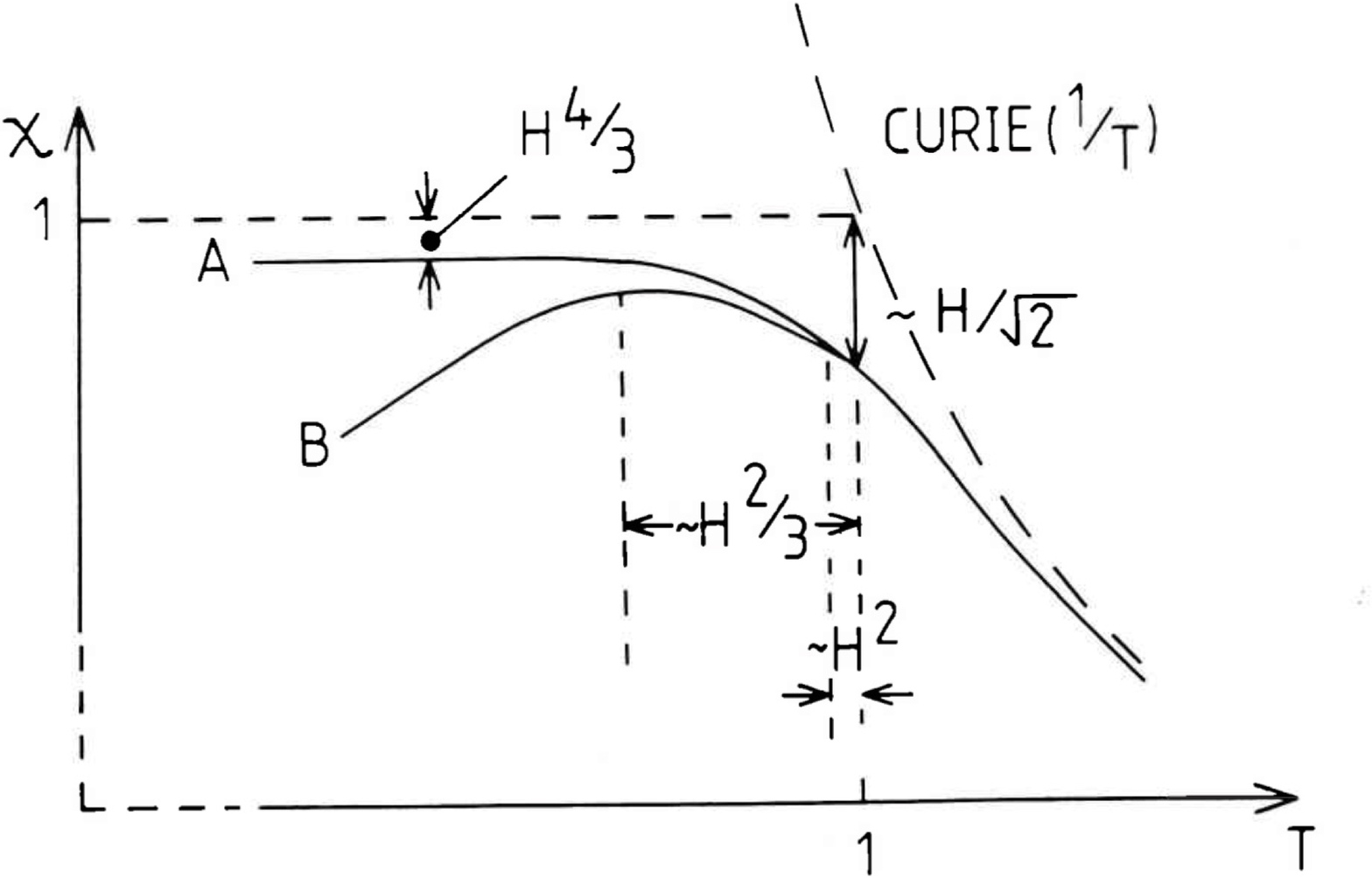}\\(b)}
 \caption{Schematic longitudinal spin glass susceptibilities in field $H$:
        (a) Ising, 
        (b) $m$-vector.
 The upper curves are field-cooled, the lower zero-field-cooled, 
 reflecting the difference between 
$[1-\int{dx q_{||}(x)}]/T$ 
and $[1-q_{||  max}]/T$. 
Figures from \cite{Sherrington1983}, with permission \copyright Physical Society of Japan(1983).}
 \label{m-vector}
 \end{figure}

{\bf{First}}, consider the application of a uniform uni-directional magnetic field, in a  direction denoted below by 1,
\begin{equation}
H= -\sum_{(ij)}{J_{ij}{\bf{S}}_{i}.{\bf{S}}_{j}} -\sum_{i} h S^{1}_{i} .
\end{equation} 
In 1981, Gabay and Toulouse \cite{Gabay1981} pointed out that this system would  now have a new transition to a spin glass phase  transverse to the field direction ({\it{i.e.}} symmetric in the $m \neq 1$ hyper-plane) beneath a characteristic (GT) line in $(h,T)$,  starting from $T=1$ for $h=0$ and with $h_{GT}$ going asymptotically to $\infty$ as $T \rightarrow 0$. Replica symmetry breaks  beneath this line, gradually and strongly  in the transverse direction (as for an ($m-1$)-dimensional spin glass) but also 
weakly in the 1-direction, until a cross-over to stronger RSB  around an AT-like $( h, T)$  line \cite{Cragg1982, Elderfield1982, Toulouse1982}; see Figs {\ref{GT}}  and   {\ref{m-vector}}.

\vspace{- 0.2 cm}

\begin{figure}[h]
\sidebyside
{
\includegraphics[width=1.8in]{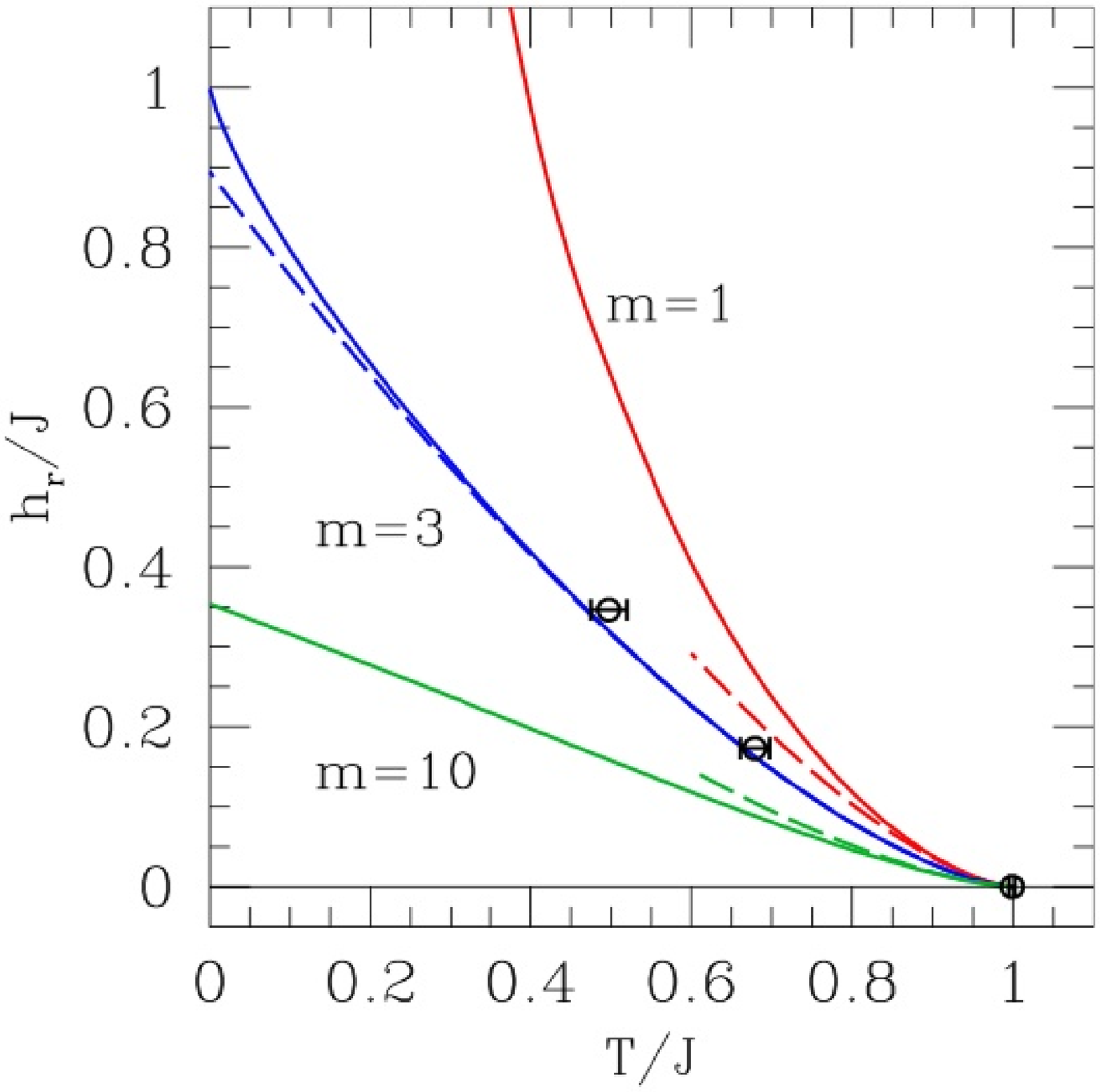}
\caption{AT line for $m$-vector spin glasses in quenched $m$-isotropic local random fields \cite{Sharma2010}, with permission. \copyright Amerlcan Physical Society (2010).}
\label{Sharma-Young}
}
{
\includegraphics[width=1.8in]{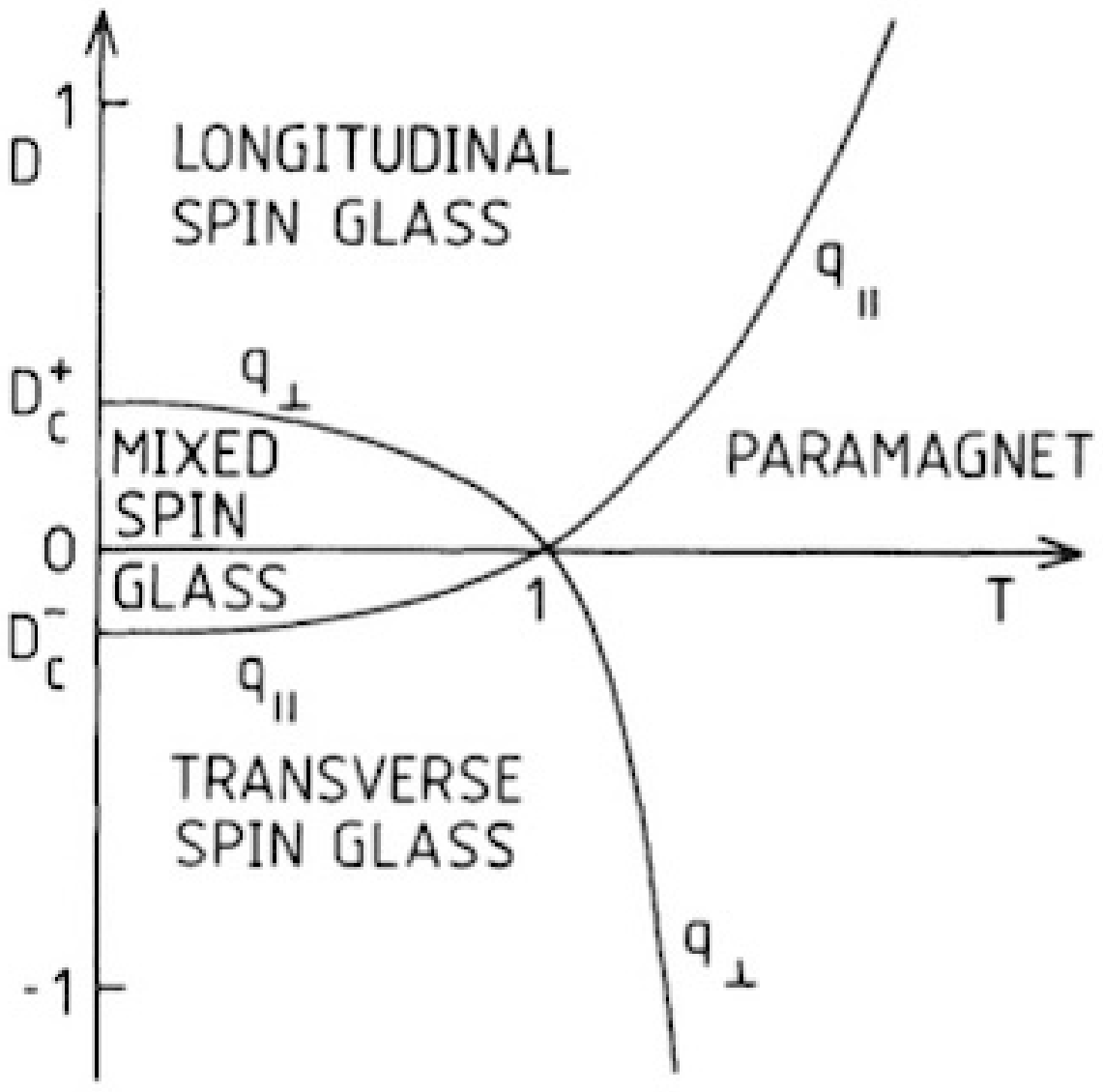}
 \caption{ Phase diagram of anisotropic spin glass. Figure from  \cite{Cragg-anisotropy} \copyright IOP Publshing; see also \cite{ Roberts-Bray}. }
 \label{CS}
 }
 \end{figure}

{\bf{Second}}, for 
a $J_0=0$  
$m>2$-vector system with external fields randomly and independently distributed 
over the full $m$-hypersphere there is no GT analogue, only AT lines which are $m$-dependent \cite{Sharma2010};
see Fig.({\ref{Sharma-Young}}).

{\bf{Third}}, consider the case with also quadratic single-site anisotropy
\begin{equation}
H= -\sum_{(ij)}{J_{ij}{\bf{S}}_{i}.{\bf{S}}_{j}} -\sum_{i} h S^{1}_{i}  -\sum_{i} D(S^{1}_{i})^2.
\end{equation}
It exhibits three different spin glass phases already for $J_0=0, h=0$,
as shown in Fig.(\ref{CS}). All three phases exhibit full RSB.  For  a more general situation, including $J_0\neq 0$, $h\neq 0$, RSB and Parisi solutions, see \cite{Elderfield_anisotropy1983}.

\subsection{Potts spins}
All $m$-vector spins, including Ising ($m=1$), have definiteness symmetry with pairwise interactions,  $J_{ij}$ and $-J_{ij}$ each leading to a unique relative ordering of the pair of spins $i$ and $j$ to minimize their energy.  This is not the case for  $p$-state Potts spins \cite{Wu1982} with $p>2$, where a ferromagnetic $J$ prefers identical Potts states on {\it{i}} and {\it{j}}, whereas an antiferromagnetic $J$ leads to $(p-1)$ degenerate optima. To investigate possible consequences of the lack of this symmetry, Elderfield and Sherrington (1983) \cite{Elderfield-Potts1983} considered the $p$-state Potts extension of SK\footnote{ Erzan and Lage \cite{Erzan1983} also introduced and studied the model, contemporaneously but independently.}
\begin{equation}
H={\sum_{(ij)}   J_{ij}   (p{\delta_{\pi(i)\pi(j)}}    -1) }  ; \pi=1,..p
\end{equation}
where the $J_{ij}$ are distributed as in eqn.({\ref{SK}}).

\begin{figure}[ht]
\centering
\includegraphics[width=1in, height=0.5 in]{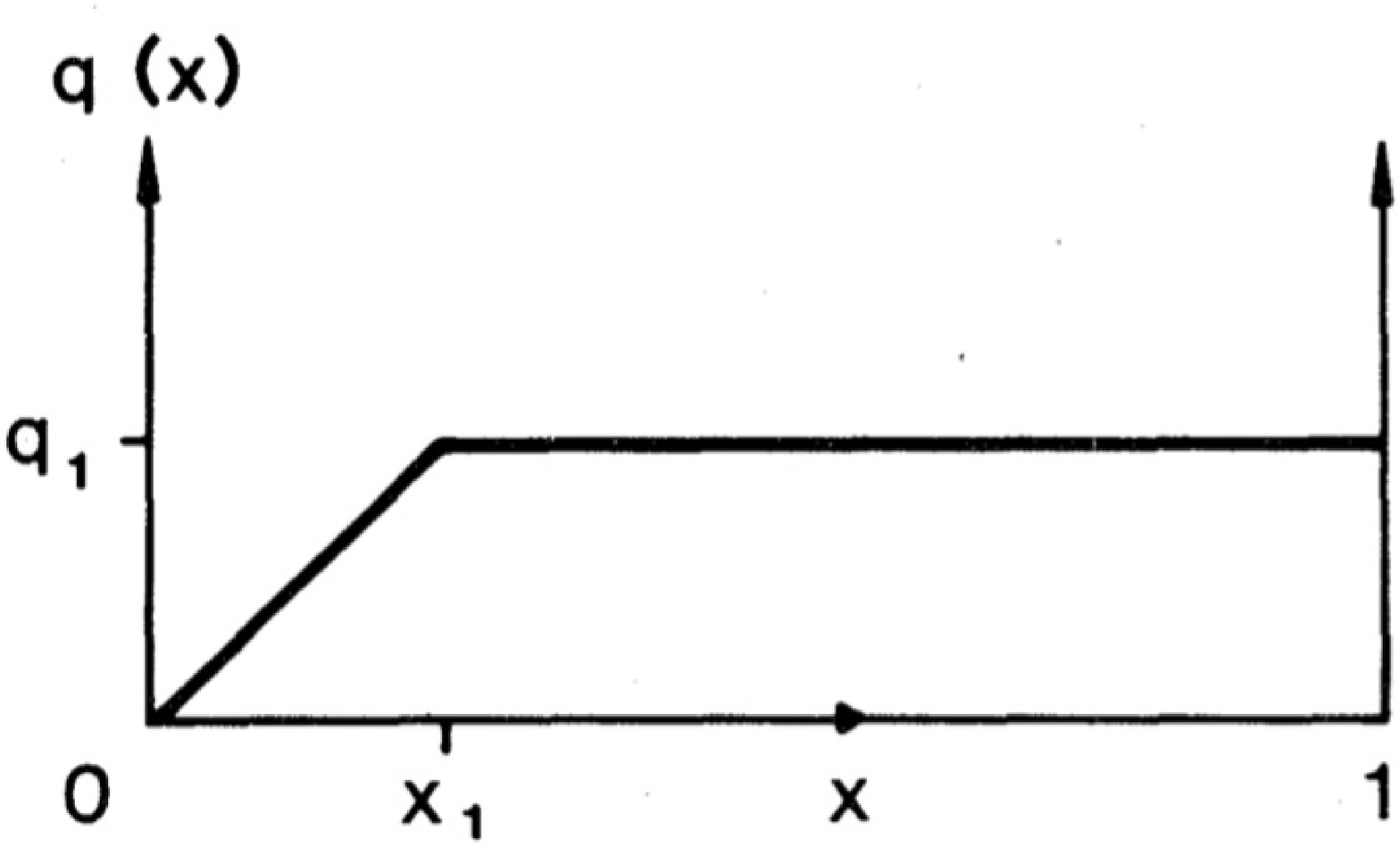} 
\includegraphics[width=1in]{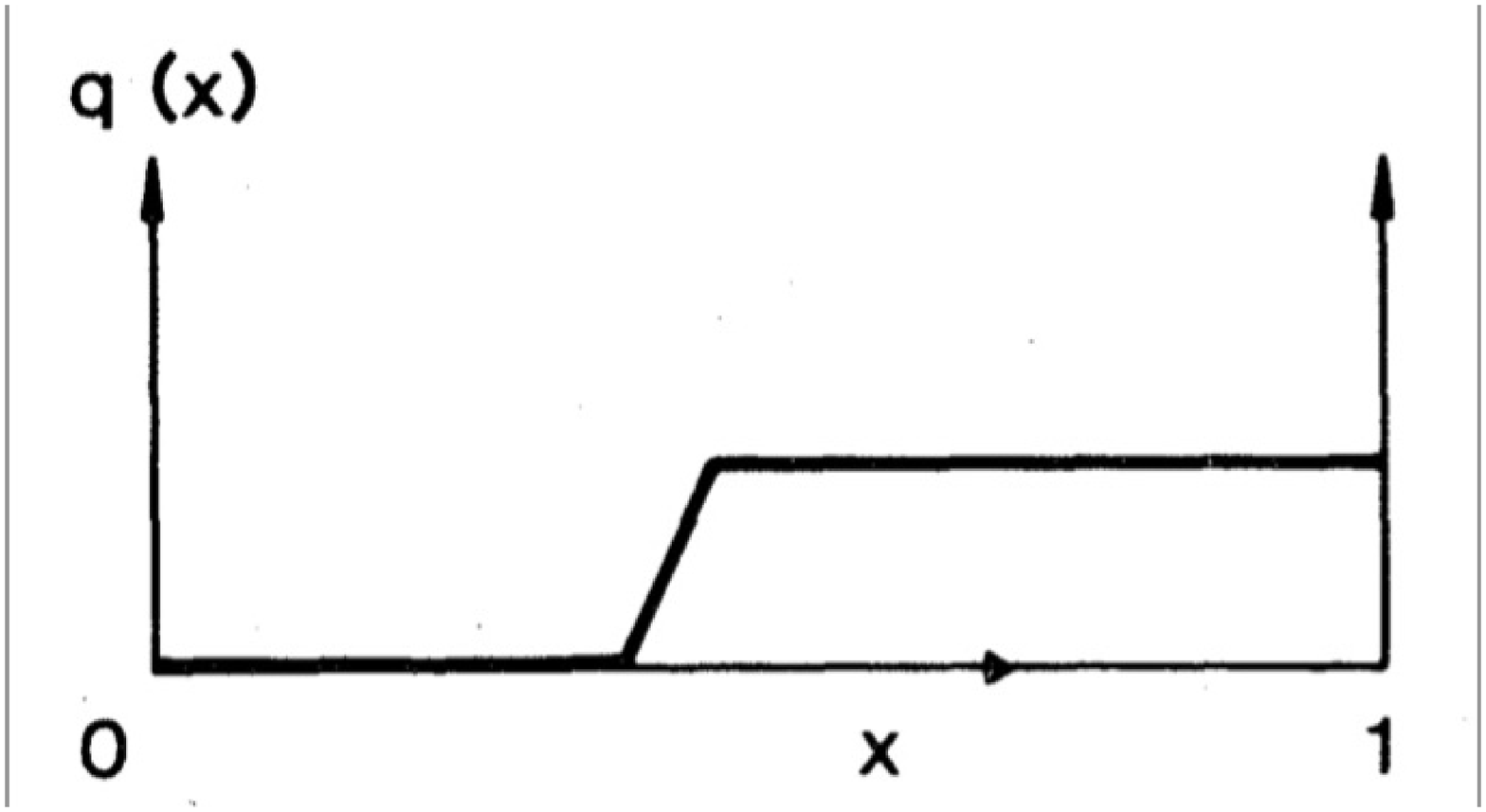}
\includegraphics[width=1in]{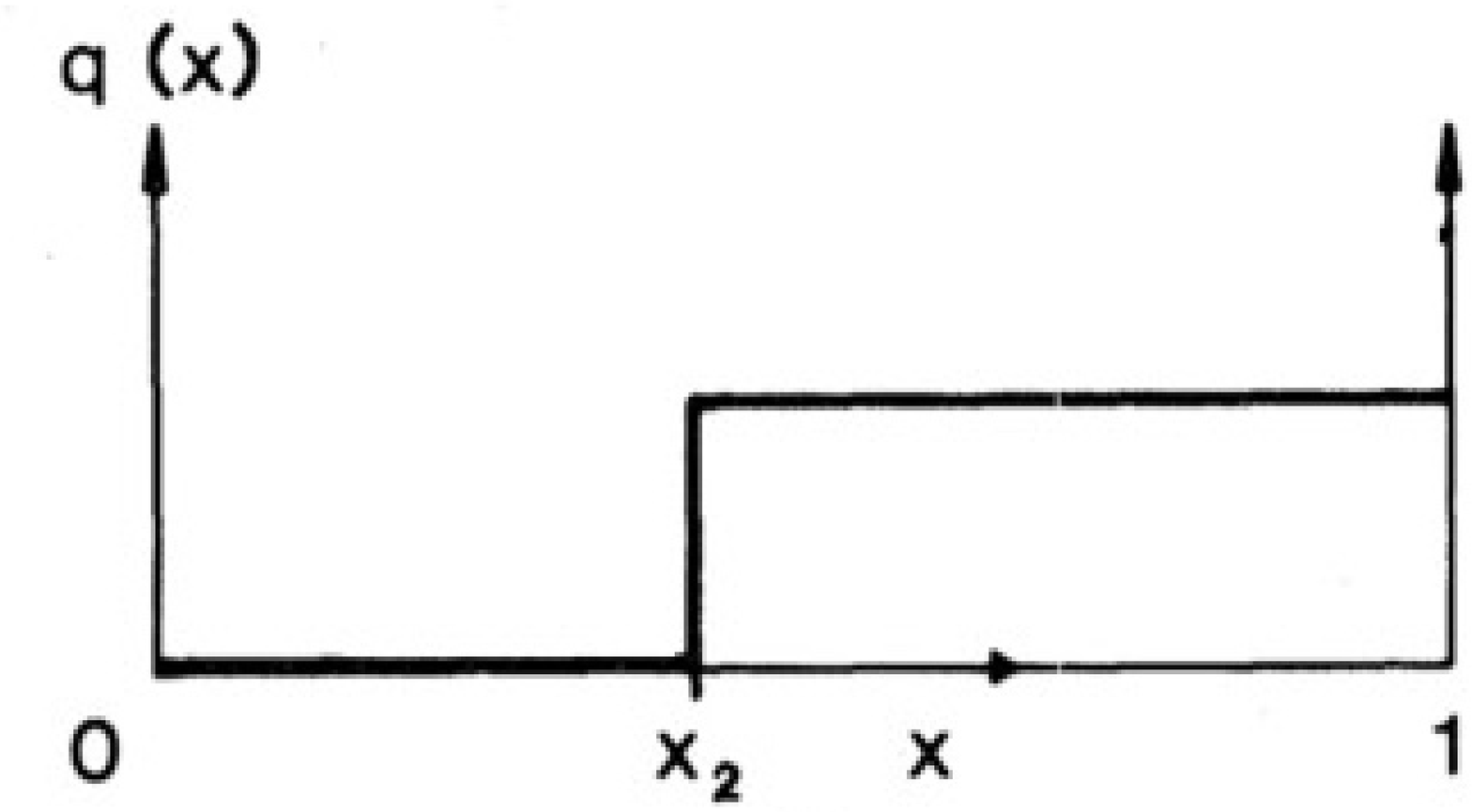}
\includegraphics[width=1 in, height=0.5 in]{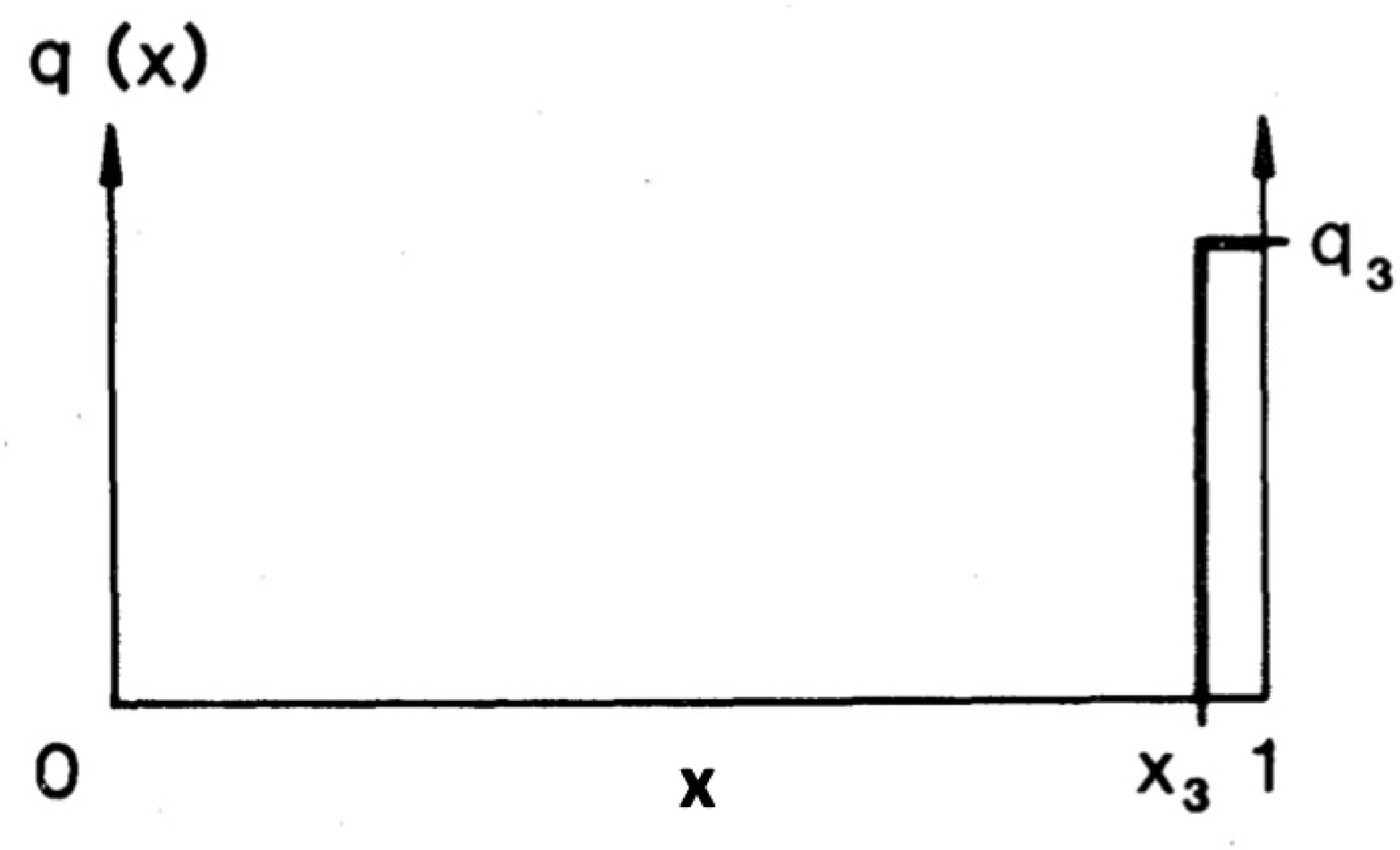}
 \caption{Potts Glasses: Parisi order functions 
 for small $\tau = (T_c - T)/T_c$; 
 \newline
(i) $p=2\equiv$ Ising (FRSB), (ii) $p_{c1} >p>2$, 
(iii) $p_{c2}>p> p_{c1}$ (C1RSB),
\newline
(iv) $p>p_{c2}$  (D1RSB). Figures from \cite{Sherrington1986}. \copyright Physical Society of Japan (1986)}          
 \label{Parisi_Potts}
\end{figure}

This system turned out to have unusual RSB behaviour compared with the Ising  higher $m$ vector systems \cite{Elderfield-Potts1983, Gross-Potts, Sherrington1986}, as illustrated in Fig ({\ref{Parisi_Potts}}) for the Parisi order function $q(x); x\in (0,1)$ for increasing $p$, based on formulae interpolated to real $p$.

\vspace{-0.5 cm}
 \begin{figure}[ht]
\centering
\includegraphics[width=1.45in]{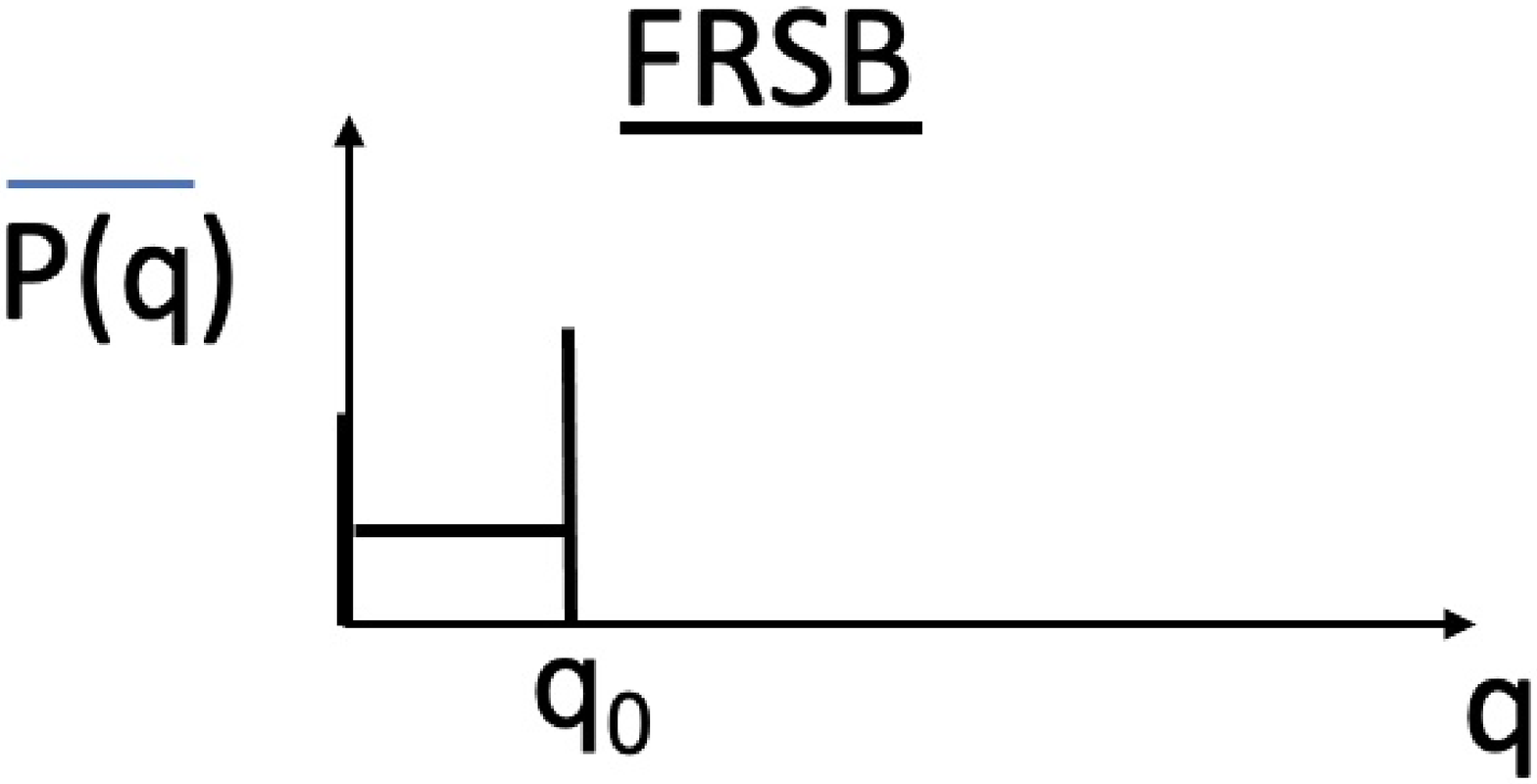}
\includegraphics[width=3.in]{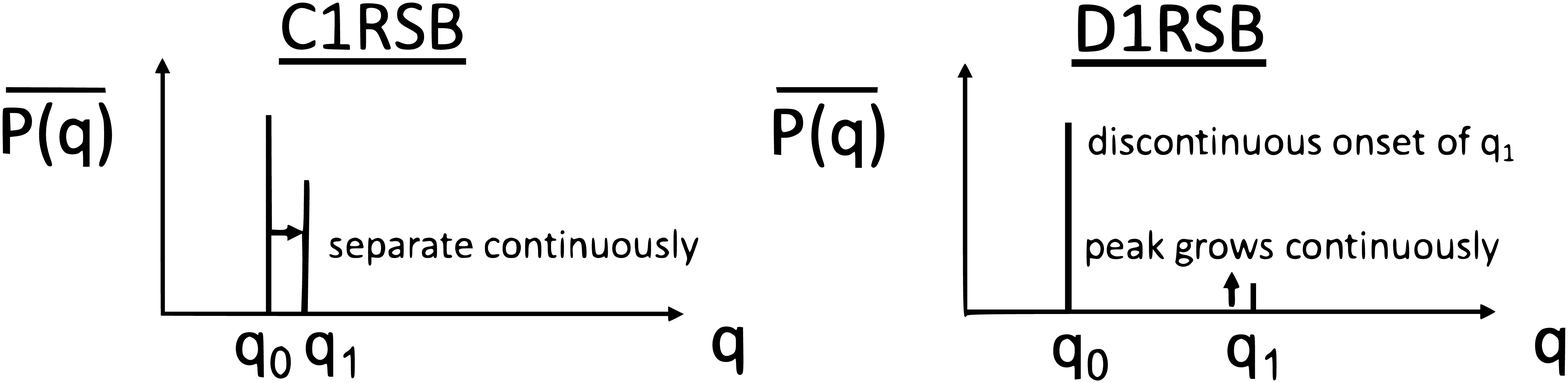}
 \caption{Types of RSB $\overline{P(q)}$: 
 (i) FRB ($p_{c1} >p>2$); (ii) C1RSB (continuous onset)
 ($p_{c2}>p> p_{c2}$) 
 (iii) D1RSB (discontinuous onset) ($p>p_{c2}$ )
 }
\label{1RSB}
\end{figure}

There are two critical Potts dimensions characterising significant changes in the overlap distribution; $p_{c1}= 2.82$ which separates a 
range of overlaps continuous between peaks at $q=0$ and a finite $q_{EA}$, {\it{i.e.}} full RSB, and another with  1RSB, itself split into two parts, the first part onsetting continuously up to the second critical value $p_{c2}=4$, the second part for $p>p_{c2}$, for which the 1RSB onset is discontinuous one-step \cite{Gross-Potts} but without latent heat, now known as a `random first order transition' (RFOT)\cite{Ted_Kirkpatrick1989} in recognition of the discontinuity in state-overlap as opposed to normal Ehrenfest first order; see  Fig. (\ref{Parisi_Potts}). Corresponding overlap distributions, $\overline{P(q)}$,
 are illustrated in Fig (\ref{1RSB}).
 
 Qualitatively similar behaviour is found in other systems lacking symmetry of definiteness; {\it{e.g.}} \cite{Goldbart-S}

  \section{{\it{p}}-spin interactions}
  
 
 Another interesting and influential extension of the original binary-interaction SK model was introduced  by  Derrida in 1980 \cite{Derrida, Derrida2}, with the binary interaction term in the SK Hamiltonian  replaced by one involving $p$ interacting 
 spins.
 
  \subsection{Ising spins}
 Derrida considered {\it{p}} interacting Ising spins,
%
\begin{equation} 
H_p= -\sum_{(i_{1} i_{2} ....i_{p})} J_{i_{1}....i_{p}}{\sigma}_{i_{1}}{\sigma}_{i_{2}}.....{\sigma}_{i_{p}}; {\hspace{0.2 cm}} \sigma =\pm 1
 \end{equation}
 with the $J_{i_{1}....i_{p}}$ randomly chosen from a Gaussian  probability distribution of variance $J^{2}/N^{(p-1)}$.
For $p\to \infty$ 
he
showed that the problem simplifies into a random energy model (REM) which he solved exactly, demonstrating freezing into the lowest energy state below a critical temperature, corresponding to 1RSB but without the need for a replica formulation.

 Subsequently, in 1984, Gross and M{\'e}zard \cite{Gross-Mezard_1984} further considered the $p\to \infty$ model with several of the different methods used for the SK model -- including the replica method -- to explicitly exhibit the structure of the spin glass phase as 1RSB with a discontinuous transition. 

In a further important study in 1985  \cite{Gardner1985}, Gardner demonstrated for $p$-spin Ising systems that (i) for all $p>2$ the spin glass transition from paramagnet to spin glass is 1RSB (whereas for p=2 (SK) the transition is full RSB (FRSB)) and (ii) that for $2<p< \infty$ there is also a lower temperature (Gardner) transition to FRSB. Both observations have had further important ramifications, discussed in later chapters.
 \subsection{Spherical models}
 A solvable extension of the normal ferromagnetic Ising model, known as the {\it{spherical model}},  was introduced by  \cite{Berlin-Kac}, replacing the Ising variables $\sigma=\pm{1}$ on each site 
$i =1..N$
 by a single (hyperspherical) condition $\sum_{i}^{N}{\sigma_{i}}^2 =N$, with the individual $\sigma$ otherwise able to take any real value.
A corresponding spherical extension of the (infinite-range $p=2$)  SK Ising spin glass model was introduced and solved 
by Kosterlitz et al. \cite{Kosterlitz1976}, shortly after SK, using several different mathematical procedures (including without needing replicas), and shown to exhibit a spin glass phase without replica symmetry breaking.

In a later study of the equilibrium statistical mechanics (statics) for the case of a spherical extension of the general-$p$-spin infinite-range (SK-extended) model, Crisant and Sommers (1992) showed that for $p>2$ the spin glass state is 
1RSB, with onset discontinuous below a critical magnetic field, continuous above it \cite{Crisanti-Sommers} . In a further (1993) paper on the relaxational dynamics with Horner  \cite{Crisanti-Sommers-Horner} they demonstrated several further intriguing features, including that the onset temperatures for 1RSB is higher for dynamics than that given by equilibrium statics. Later work recognised that the dynamical 1RSB onset  corresponds to marginal stability of the statics. 

Also in 1993, Cugliandolo and Kurchan \cite{Cugliandolo-Kurchan} made a different study of the dynamics of the $p$-spin spherical model at long but finite times, demonstrating many unexpected but important new features, including `weak' ergodicity breaking and aging effects, leading to much further activity and results, which will be expanded upon in later chapters. 

Even though its statics do not exhibit RSB, the $p=2$ spherical model was shown also to have non-trivial dynamics \cite{Cugliandolo-Dean}.

\begin{figure}[ht]
\centering
{
\includegraphics[width=1.2in]{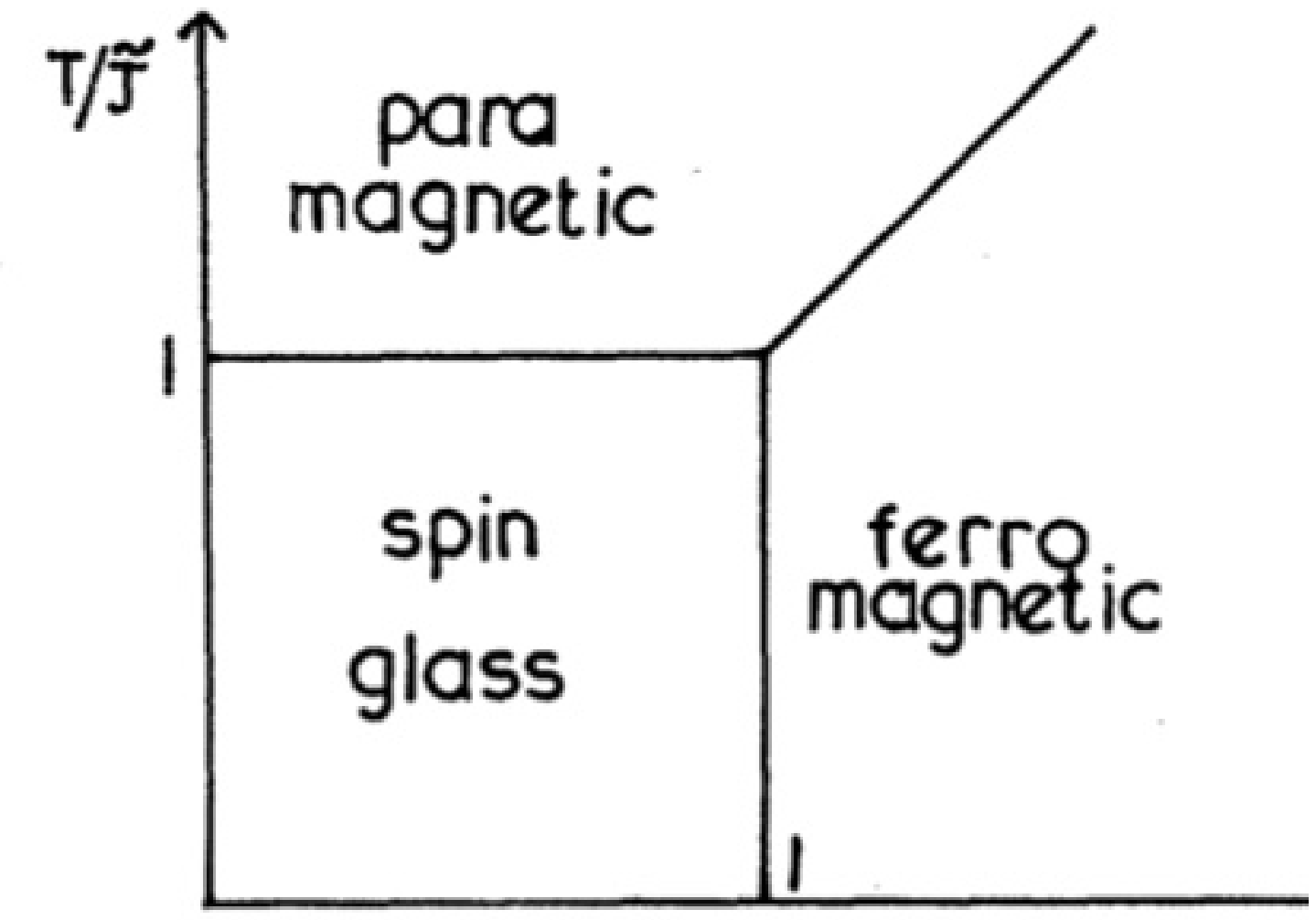}
 \label{p=2_spherical model}
 }
{ 
\includegraphics[width=1.5in]{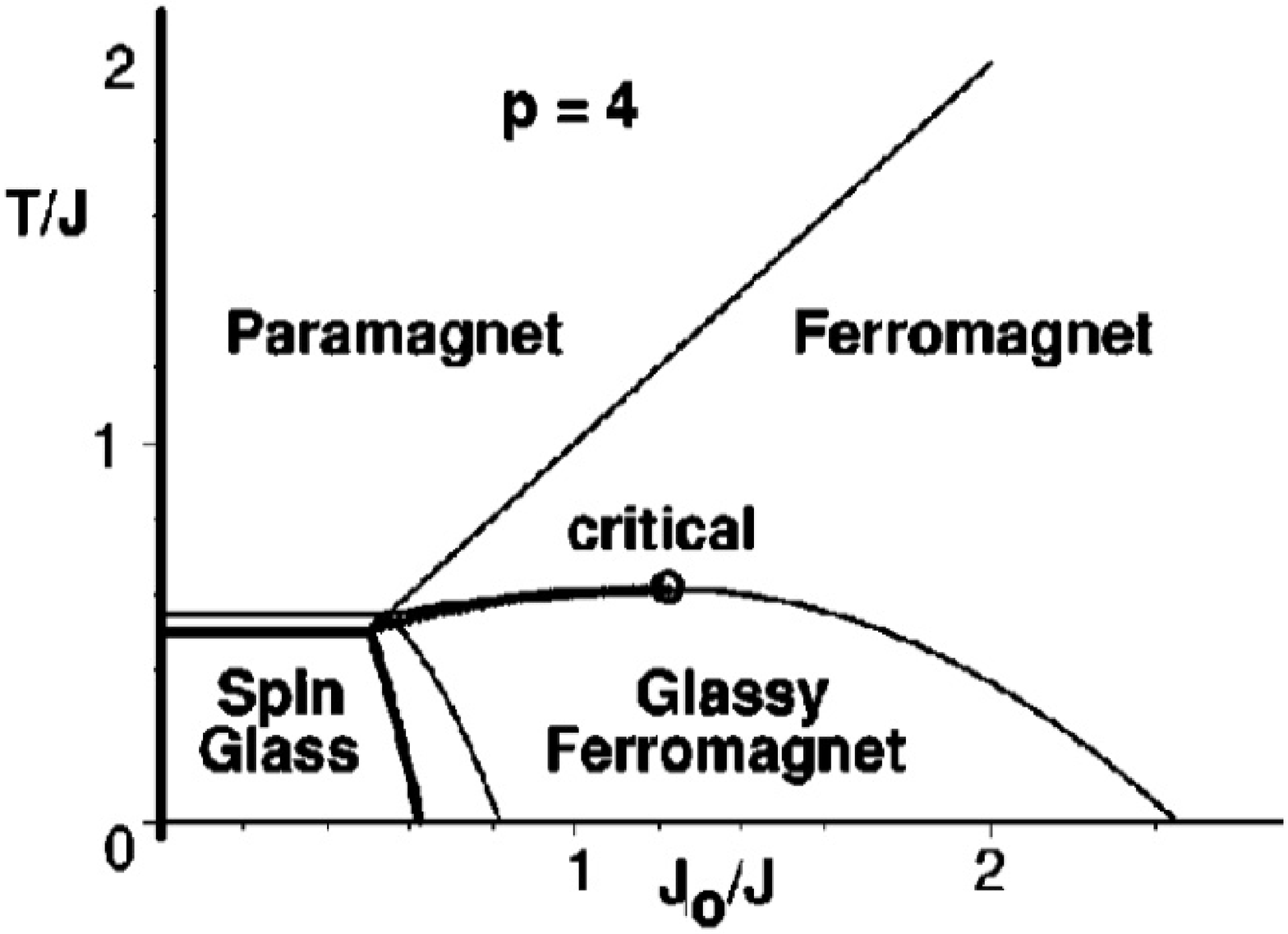}
 \label{Hertz_p=3}
}
 {
 \includegraphics[width=1.5in]{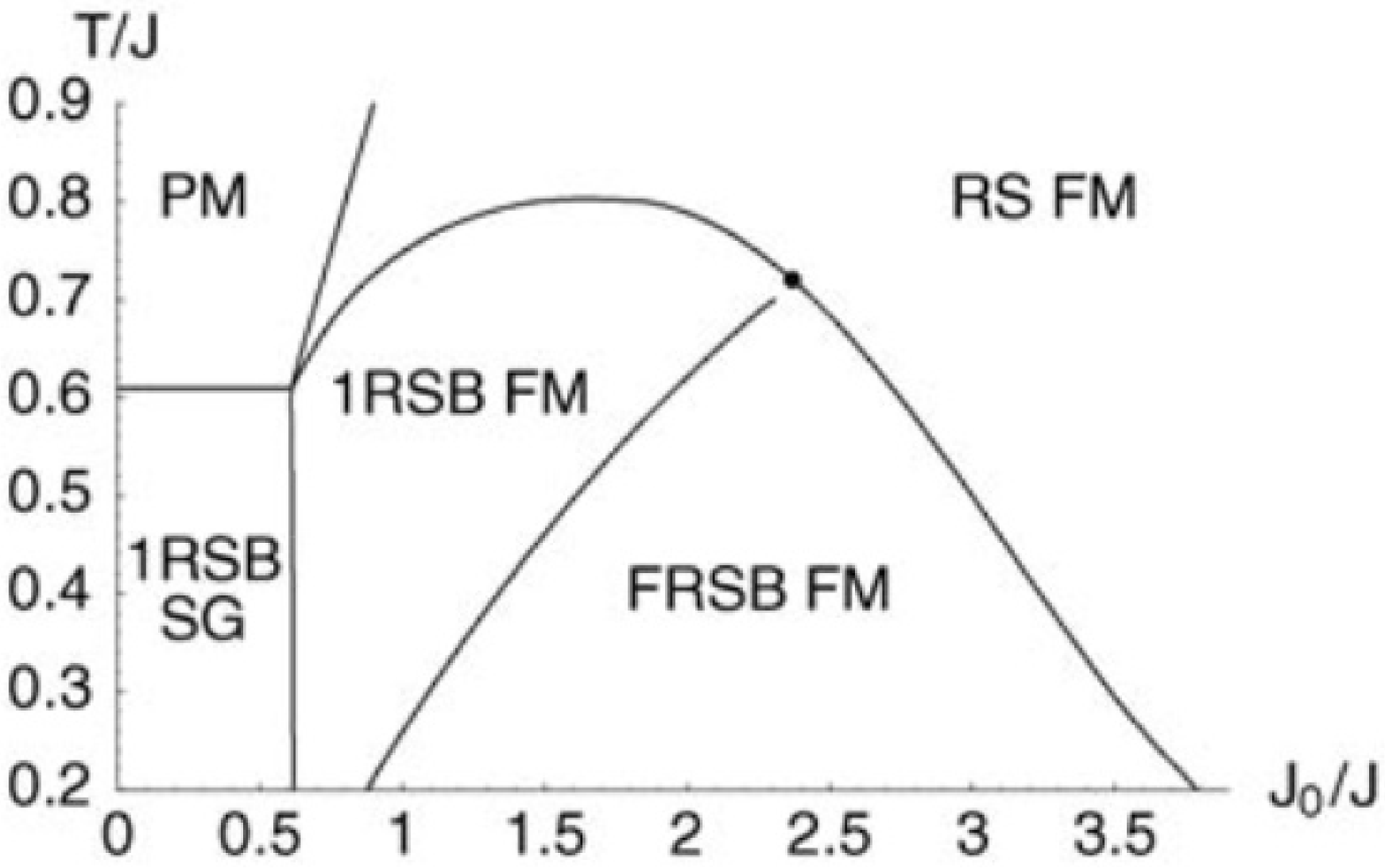}
 \label{p=5,r=2}
 }
  \caption{Phase diagrams for $p$-spin random zero-mean interactions plus 2-spin ferromagnetic interactions:
  (i) spherical spins  $p=2$ (from \cite{Kosterlitz1976} with permission. \copyright American Physical Society (1976)), statics, no RSB;
 (ii) spherical spins $p=4$, statics (solid lines), dynamics (fine lines), dot (critical point, discontinuous 1RSB to left,  continuous 1RSB to right.) (from \cite{Hertz1999} with permission. \copyright American Physical Society (1999) ); (iii) Ising $p=5$ statics, showing also 1RSB to FRSB transition (from \cite{Gillin-Nishimori} \copyright  IOP Publishing (2001).}
 \label{SG_phase_diagrams}
 \end{figure}

Fig. (\ref{SG_phase_diagrams}) shows examples of several models with various $p$-spin random zero-mean SK-extended interactions together with 2-spin ferromagnetic interactions, demonstrating several phases, including glassy ferromagnetism, and transitions.

 There has also been significant interest in mixed-$p$ spin glasses with infinite-ranged  extended-SK interactions, particularly spherical models  because of their solvability advantages along with  realisation that they (mixed-$p$ systems) can have FRSB as well as 1RSB, along with more complicated issues, both static and dynamic, of potential relevance to the understanding of other glasses; see {\it{e.g.}} \cite{Crisanti-Leuzzi, Crisanti-Leuzzi2007, Franz2020} and later chapters in this book.

 \section{Finite-connectivity random graph models}
Although the infinite-ranged SK model was introduced for its putative solvability rather than to mimic specific experimental systems, subsequent interest (and extension of the model) turned to many other problems where spatial separation is irrelevant, e.g., communication through the world wide web and random satisfiability in computer science. Several such problems are discussed in later chapters, but we here (briefly) discuss a few
which appeared relatively early on in the field.

In 1985, Viana and Bray introduced a dimension-free but (on-average) finite-connectivity alternative to the SK spin glass model [38], 
in which active spins are effectively located on an Erd\"os-Renyi graph of average connectivity $c$ with interactions on linked sites quench-randomly chosen from a distribution $P_{c}(J)$:
 \begin{equation}
 H= -\sum_{(ij)}  J_{(ij)} {{\sigma}}_{i}{{\sigma}}_{j} ; 
 \hspace{.5cm}  
 P(J_{ij}) = (c/N)P_{c}(J_{ij})+(1-c/N)\delta(J_{ij}).
 \end{equation}
 As for EA and SK, a spin glass phase results from sufficient frustration in $P_{c}(J)$, in purest form with
 $P_{c}(J)=P_{c}(-J)$, but without need for $N$-scaling of the mean or variance for $c$ finite.
 
In 1986 Fu and Anderson  \cite{Fu-Anderson} showed its relationship to graph bipartitioning\footnote{In the optimal graph bipartioning problem one seeks a separation of vertices into two groups of equal size with the minimal number of graph edges between them.This problem is equivalent to finding the ground state of of a system of  Ising spins on the vertices of the graph interacting ferromagnetically along the edges but with a frustrating constraint of zero overall magnetization.}, including a demonstration of mapping the extensively connected graph problem to the SK model.
 
The Erd\"os-Renyi graph above has variable connectivity on different sites with only its average given by $c$. However, there is 
also much interest  --  particularly in computer science -- 
in what are known as random {\it{regular}} graphs, where all vertices have same connectivity $c$,  but still quench-disordered. This problem, with finite $c$, independent of the total number of vertices $N$, was considered numerically by Banavar {\it{et al.}}
\cite{Banavar1987} in connection with optimal graph bi-partitioning, 
finding evidence for several of the novel features of the Parisi SK solution. The full analytical solution has proven harder than for SK, but RSB was found  \cite{Wong1988, Franz2001, Mezard-Parisi2003}. 
Further finite-connectivity random graph optimization studies, and extensions to classic computer science problems such as random satisfiability, have led to many new concepts,  some of which will be discussed in later chapters.

 \section{Induced moment spin glasses}
 
 Models considered thus far have  
 moments (non-zero spins) even in the absence of interactions. For some systems, however, interactions are required to bootstrap-induce moments and exhibit cooperative 
 magnetism. Ferromagnetic examples have been known for centuries and recognised for many decades, but here we make the case for spin glasses.

\subsection{Discrete spins}
 
\begin{SCfigure}
\centering
\caption{Phase diagram of Ghatak-Sherrington model; figure  from \cite{MS} \copyright IOP Publishing (1985) : 
The solid curve indicates a continuous transition between paramagnet and spin glass, the dashed curve a thermodynamically first order transition.  
The spin glass phase is full RSB.
}
\vspace{-0.1 in}
\includegraphics[width=1.9in, height=1.4in]{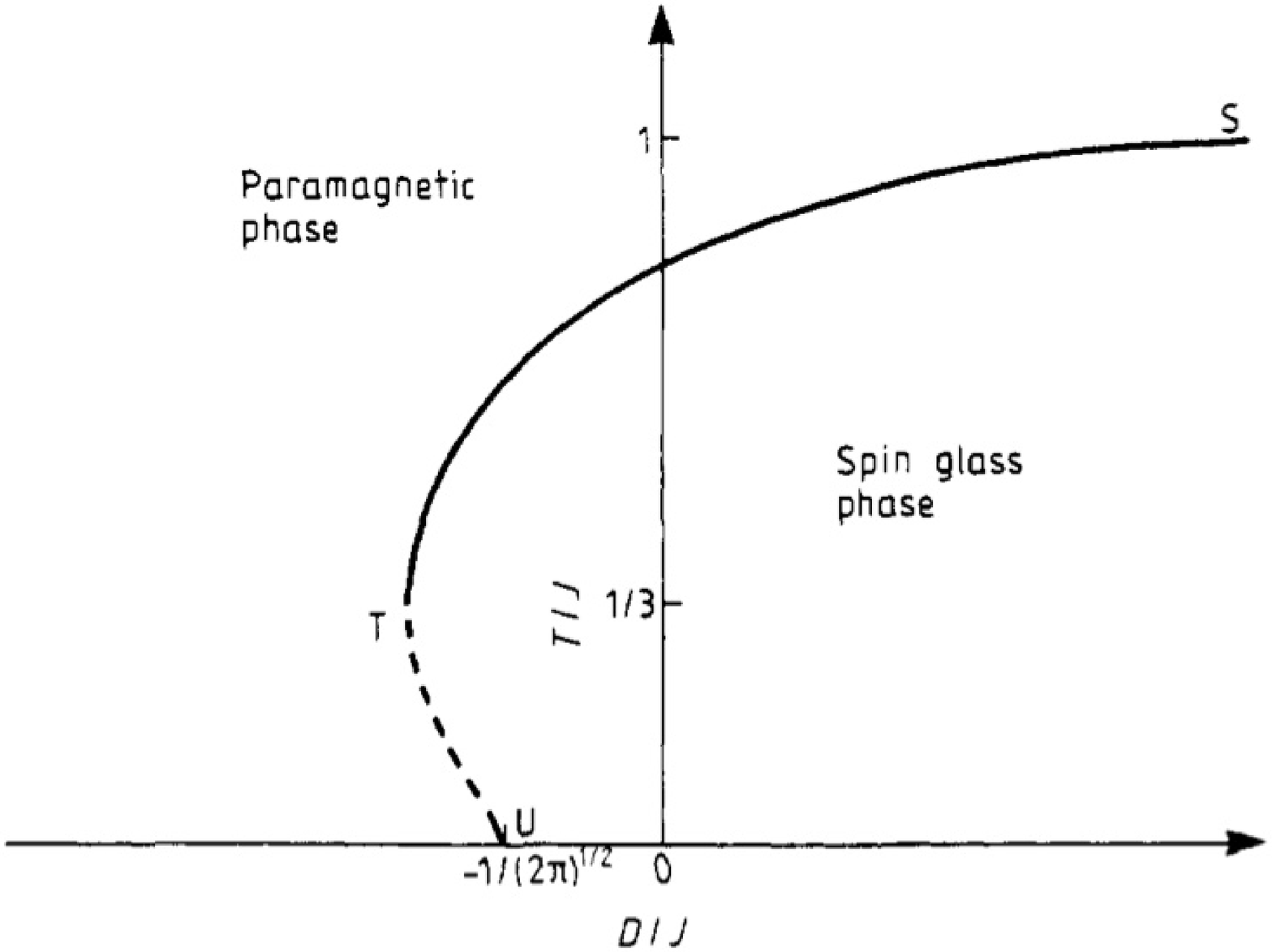}
 \label{GS}
\end{SCfigure}

 Stimulated by consideration of local crystal field splitting favouring a singlet ground state, 
 a discrete/`hard'-spin example
 was introduced in 1977 by Ghatak and Sherrington (GS) \cite{Ghatak1977}, with Hamiltonian
 \begin{equation}
H= - D\sum_{i} S_{i}^{2} -\sum_{ij}J_{ij} S_{i}S_{j}    ;  \  S=0,\pm1,
 \label{Ghatak-Sherrington}
 \end{equation}
 with the $J_{ij}$ Gaussian-distributed around zero mean. In the absence of the interaction term, for $D>0$ the local term alone favours $S=\pm 1$, but for $D<0$ it favours all $S=0$, no moments.  Consequently, for $D>0$ the full system (including the interaction term) behaves qualitatively analogously to the SK model, while for $D<0$ there is  a critical maximum $|D|$ for interactions to induce bootstrap-induced cooperative order. 
 At the RS level of description,
 the phase diagram is as shown in Fig.(\ref{GS}) and demonstrates both continuous and (Ehrenfest) first order 
phase transitions from paramagnet to spin glass, along with re-entrance/inverse-freezing from spin glass to paramagnet. Including RSB modifies the details of the first order transition line but retains its qualitative character \cite{Crisanti-Leuzzi2005}.

 \subsection{Itinerant spin glasses}
A simple {`soft'-spin induced-moment spin glass model was initially proposed in 1973 by Sherrington and  Mihill  \cite{Mihill1974, Sherrington-Moscow}, to  try to explain the magnetic behaviour of the transition-metal spin-glass alloy {\bf{Rh}}Co; see Fig(\ref{RhCo_phase_diagram}) \cite{Coles1974}\footnote{Pure bulk Rh is non-magnetic, pure bulk Co is an itinerant ferromagnet, but isolated Co ions in Rh do not carry a long-lived moment and at very low concentrations of Co {\bf{Rh}}Co is Pauli paramagnetic.  However, at intermediate concentrations a spin glass was observed, along with magnetic clustering features.}. Starting from a disordered Hubbard model alloy of ions with different Coulomb $U$, transforming to auxiliary local magnetization fields and with some simplification, including static approximation, leads to an effective Hamiltonian of the form
\begin{equation}
H= \sum_{i}   \{ {\kappa}_{i} |{\bf{m}}_{i}|  ^2
 + \lambda_{i} |{\bf{m}}_{i}|^4\} -\sum_{(ij);i \neq j} J({\bf{R}}_{ij}) {\bf{m}}_{i}. {\bf{m}}_{j},
\label{S-Mihill}
\end{equation}
where the $i,j$ label sites, the $\bf{m_i}$ are unconstrained local magnetization variables, the local coefficients $\kappa_{i}$ and $\lambda_{i}$  depend on the type of atom at site $i$, and the $J({\bf{R}}_{ij})$ are inter-site interaction energies\footnote{In terms of the Hubbard model parameters, $\kappa_{i} =(1-U_{i} \chi_{ii})$ and $J({\bf{R}}_{ij}) = (U_{i}U_{j}\chi_{ij})$, where the $U$ are the appropriate (site-occupation dependent) Hubbard potentials and $\chi_{ij}$ is the two-site conduction-band susceptibility.}.

$\kappa<0$ would favour local moments (ground state with $m \neq 0$ in the absence of interactions), the analogue of $D>0$ in the GS model; {\it{cf}}\cite{Anderson1961}.  But for both Rh and Co the $\kappa$ are positive, so that the local harmonic terms alone favour no moment, $|{\bf{m}}|=0$. Inclusion of the interaction term, however, offers the potential for bootstrapped collective order if the resultant binding energy can overcome the local cost.
\begin{figure}[ht]
\sidebyside
{
\includegraphics[width=2.2in]{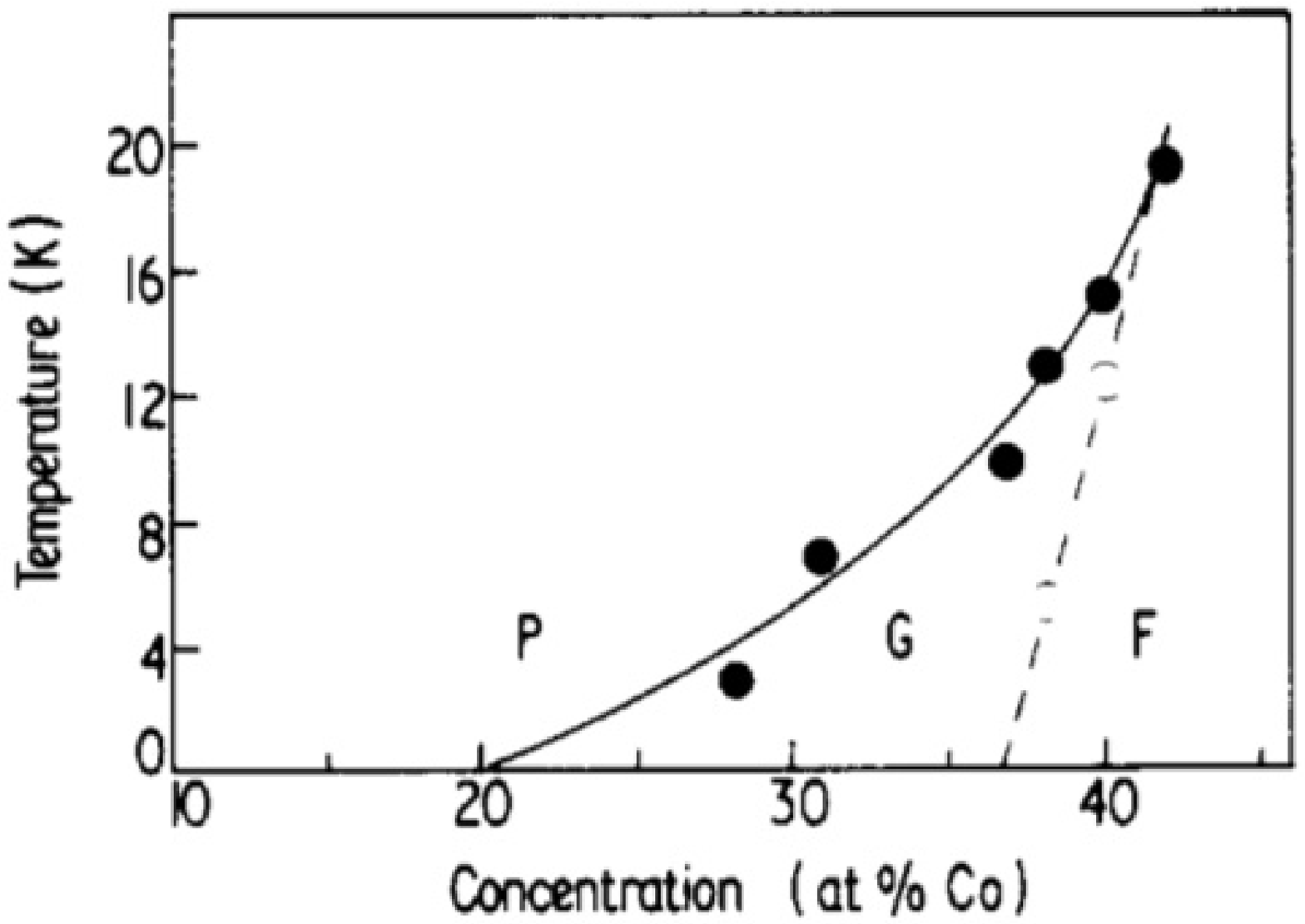}
 \caption{{\bf{Rh}}Co;
 {\bf{P}}aramagnet, spin {\bf{G}}lass, {\bf{F}}erromagnet. 
 Figure from \cite{Jamieson1975} with permission. \copyright IOP Publishing (1975)}
 \label{RhCo_phase_diagram}
}
{
\includegraphics[width=2in]{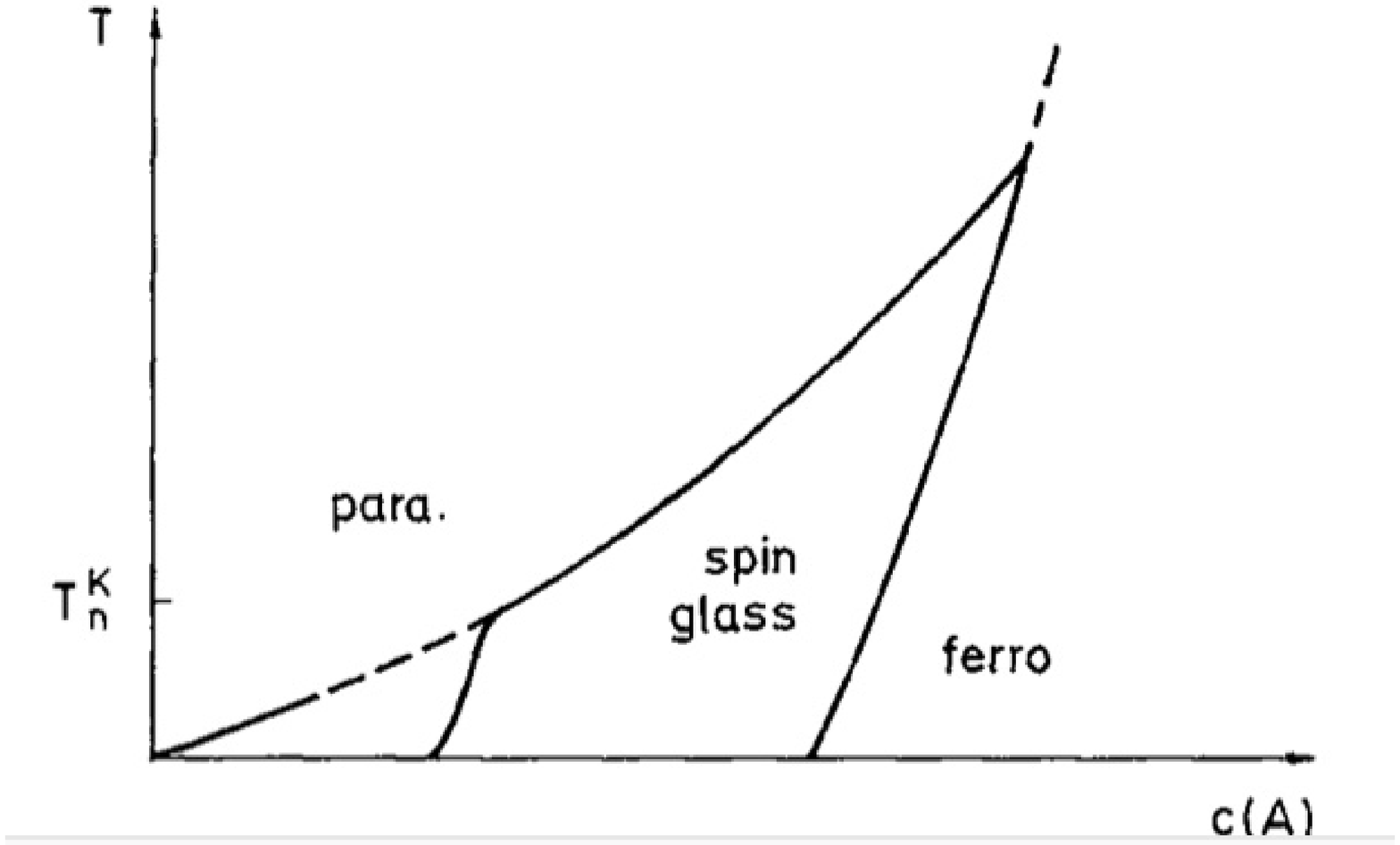}
\caption{Schematic prediction for {\bf{Rh}}Co-type alloy, from \cite{Mihill1974} with permission \copyright European Physical Journals. 
}
\label{Sherrington-Mihill}
 }
 \end{figure}
For pure Co this results in ferromagnetism, but the same is not true of Rh\footnote{ {$(1-U_{Co}\sum_{j}{\chi_{ij}})<0$},    $(1-U_{Rh}\sum_{j}{\chi_{ij}})>0$.} which consequently is not magnetic. 
Hence,
already without need for computation, 
in analogies with local moment spin glasses (such as {\bf{Au}} Fe)
 and itinerant ferromagnets (such as Ni),
 one can immediately anticipate a phase diagram
with three low temperature phases as a function of increasing Co concentration of {Pauli paramagnet, itinerant  spin glass,
and itinerant  ferromagnet; see Fig(\ref{Sherrington-Mihill})\footnote{Pre-EA, Sherrington and Mihill
\cite{Mihill1974, Sherrington-Moscow} 
 were motivated by a desire to include statistically occurring local moment-like clustering. Post-EA, Hertz  \cite{Hertz1979} examined a similar starting model, concentrating on prediction of an extended itinerant spin glass phase and proposed  naming it  {\it{Stoner glass}} in recognition of the pioneering work of Stoner \cite{Stoner} on itinerant (conduction electron) magnetism.}.

 \subsection{Relaxor ferroelectrics as `soft' pseudo-spin glasses}
 
 A similar reasoning
provides a likely explanation for a long-standing puzzle in some displacive ferroelectric alloys and also opens up a potentially interesting new direction for random-field magnets.
 
 Examples of pure displacive ferroelectrics are found in ionic crystalline compounds based on the generic formula {$\rm{ABO_3}$} with charges ${\rm{A^{2+}, B^{4+}}}$, and ${\rm{O^{2-}}}$, and at high temperature have cubic perovskite structure; see Fig ({\ref{Perovskite}}).  Inter-ionic interactions can lead to spontaneous structural distortions to lower symmetry at lower temperature if the consequential interaction energy lowering is great enough to overcome local displacement costs. With several  ionic types within the unit cell, with different harmonic displacement coefficients, the displacements of the different ion types relative to their locations in the higher-temperature phase are themselves different, leading to spontaneous intra-cell electric dipole moments and hence ferro-electricity\footnote{Displacive ferroelectricity is thus an analogue of itinerant ferromagnetism.}.  {$\rm{BaTiO_3}$ (BT) and  ${\rm{PbTiO_3}}$ (PT) are two classic examples and have been studied theoretically quantitatively using 
 non-disordered and three-dimensional analogues of eqn.(\ref{S-Mihill}), with the local magnetizations replaced by displacements; {\it{e.g}} by \cite{Zhong1995,Garcia}\footnote{First-principles quantum-mechanical calculations are used to determine the parameters in effective classical Hamiltonians whose subsequent thermodynamical consequences are then explored through computer statistical mechanical simulations.}; see again Fig ({\ref{Perovskite}}). 
 
 \begin{SCfigure}
\centering
\includegraphics[width=1.7 in, height=2.00 in]{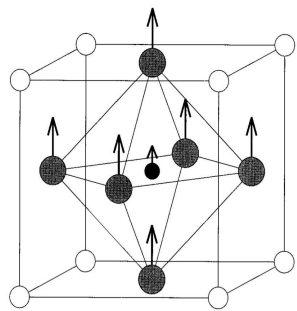}
 \label{Perovskite}
\caption[width=6.00 in]{{$\rm{ABO_3}$} perovskite. \newline 
Circles show high T cubic cell:
open circles = A, central black dot = B, dark shaded circles = O.
\newline
Unit cell distorts to tetragonal ($c>a=b$) beneath $T_c$.
Arrows show ionic displacements in the ferroelectric phase relative to the tetragonal cell.
${\rm{PbTiO_3}}$,  from \cite{Garcia} with permission \copyright American Physical Society (1996).
\newline
\newline
}
 \end{SCfigure}
 
 The present interest, however, concerns substitutional alloys.
 
Already in the 1950s, experimental behaviour  \cite{Smolenskii} retrospectively reminiscent of corresponding behaviour seen in the 1970s and 80s in experimental spin glasses \cite{Cannella-Mydosh, Mydosh-Mulder} was observed in the alloy
 $\rm{Pb(Mg_{1/3}Nb_{2/3})O_3}$ (PMN),
with the Mg and Nb ions randomly distributed on the B-sites of the {$\rm{ABO_3}$} lattice, 
;  high, sharpish, strongly frequency-dependent peaks in AC susceptibility,
 see Figs. ({\ref{AC_susc}}(i) and (ii),
 but with no change in overall lattice structure and no ferroelectricity,
 again reminiscent of early spin glass experimentation that demonstrated spin freezing
 without periodicity.
 Other characteristic spin glass features, such as differences between FC and ZFC static susceptibilities beneath the static/low frequency susceptibility peak temperature, were observed later; see {\it{e.g.}} Fig. (\ref{FC_ZFC}).  Initially referred to simply as a {\it{ferroelectric with diffuse phase transitions}}, later the name {\it{relaxor}} was attributed to the new phase.
Given their experimental similarity, theoretical comparison with spin glasses, with induced electric dipoles the analogues of local magnetizations, seems natural \cite{Viehland, Pirc-Blinc, Akbarzadeh2012, Sherrington2014}.

 \begin{figure}[ht]
\centering
{
\includegraphics[width=1.0in, height =1.3in]{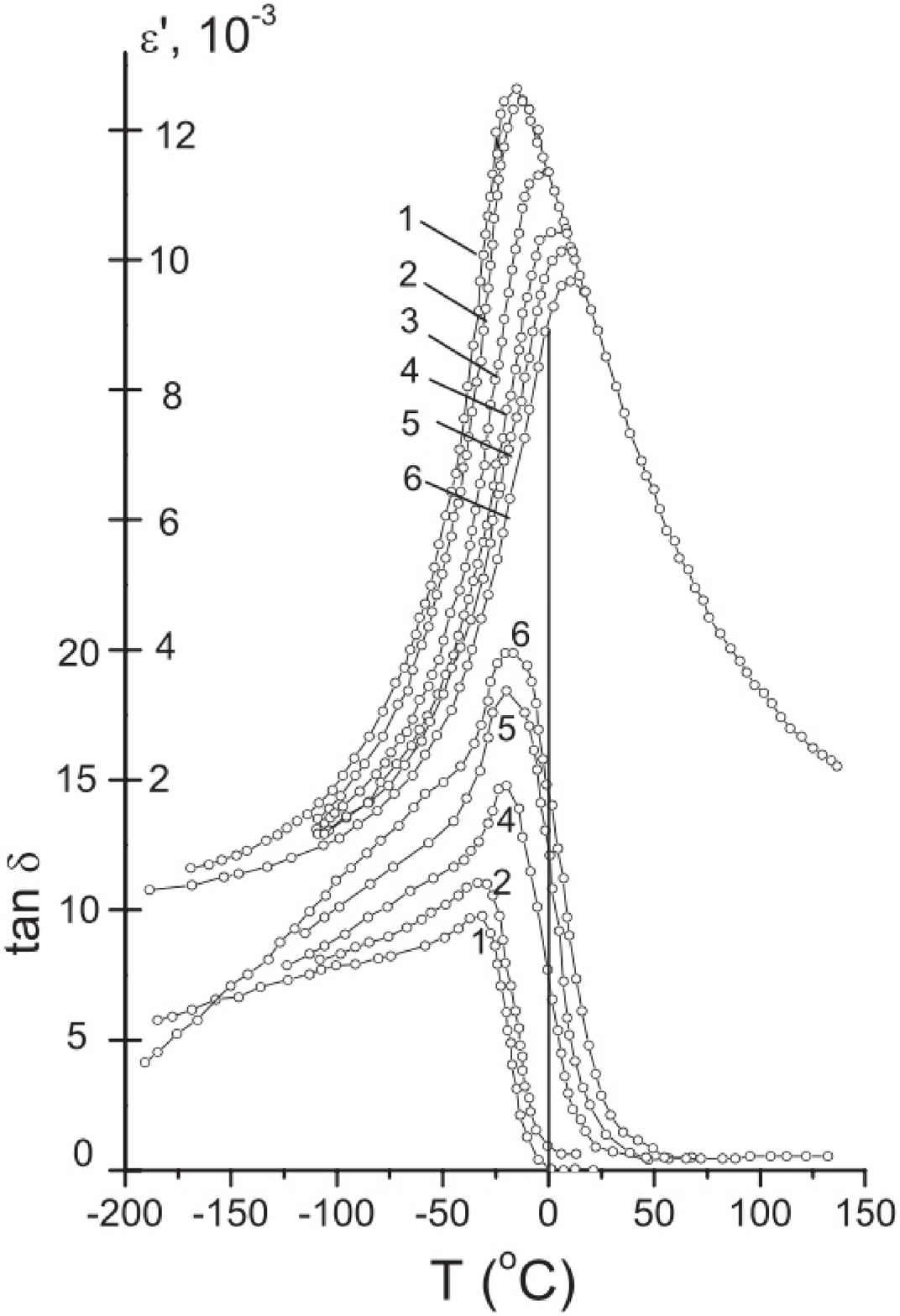}
 }
{ 
\includegraphics[width=1.0in, height=1.3in]{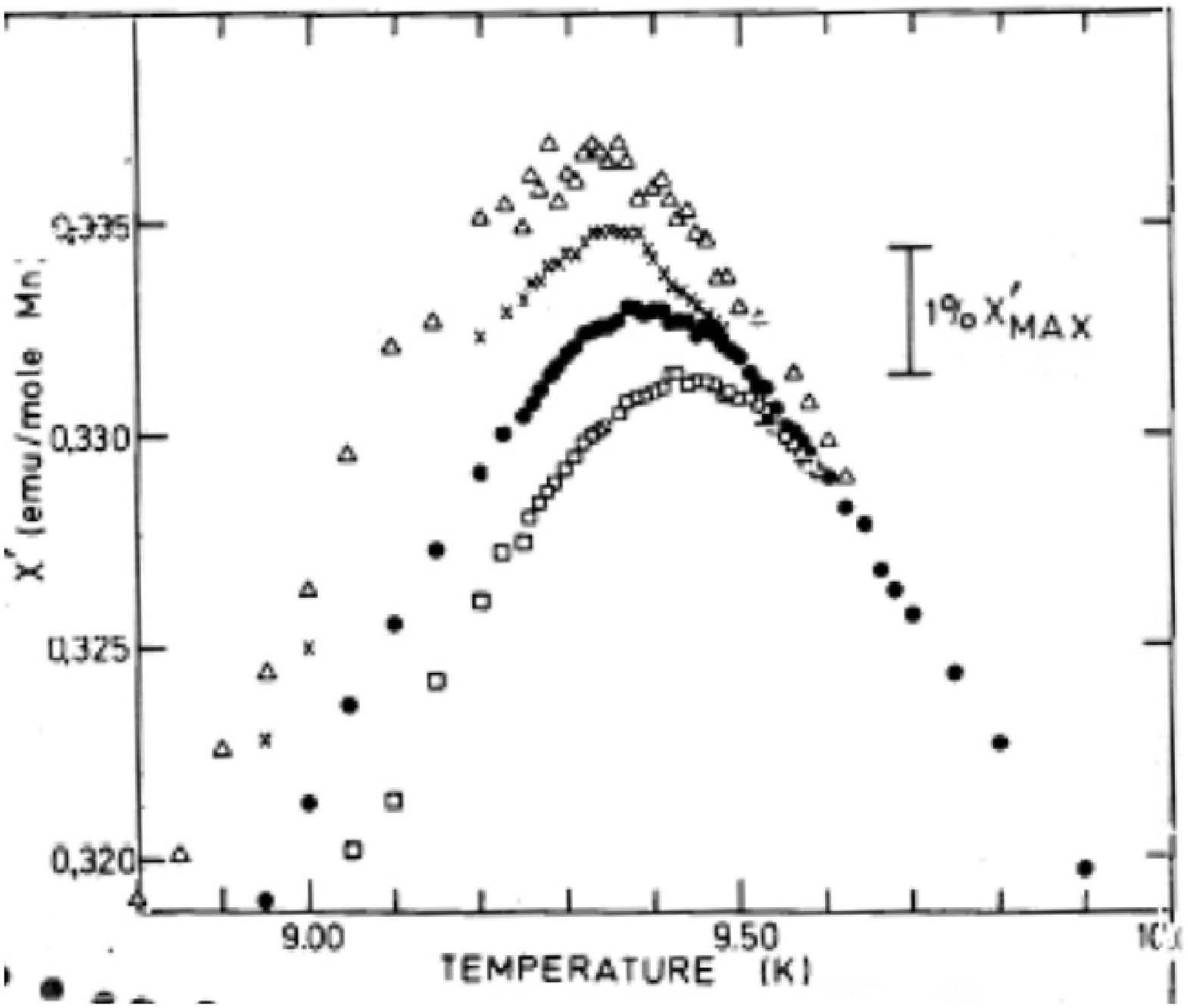}
}
 {
 \includegraphics[width=1.0in, height=1.38in]{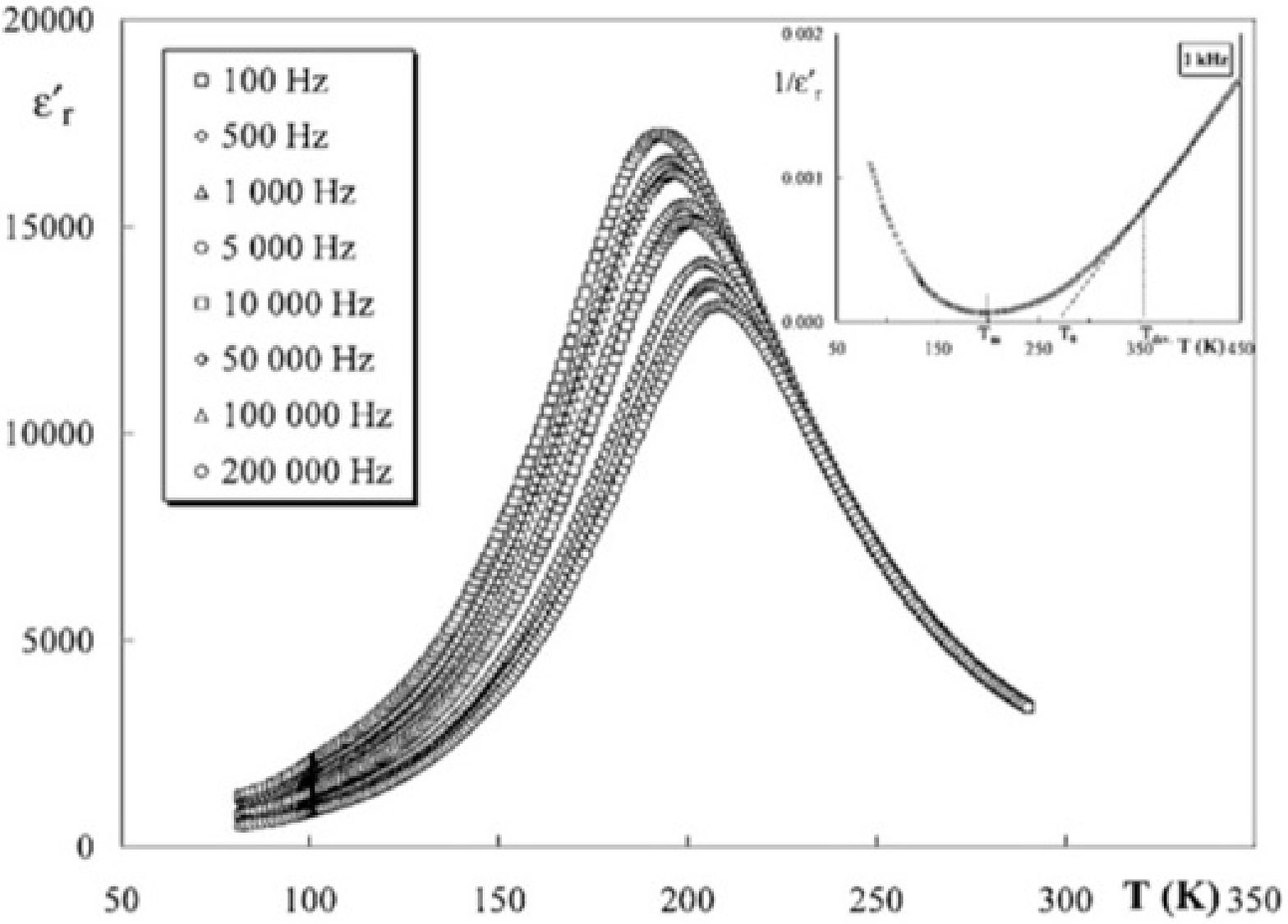}
 }
 {
 \includegraphics[width=1.1in, height =1.3in]{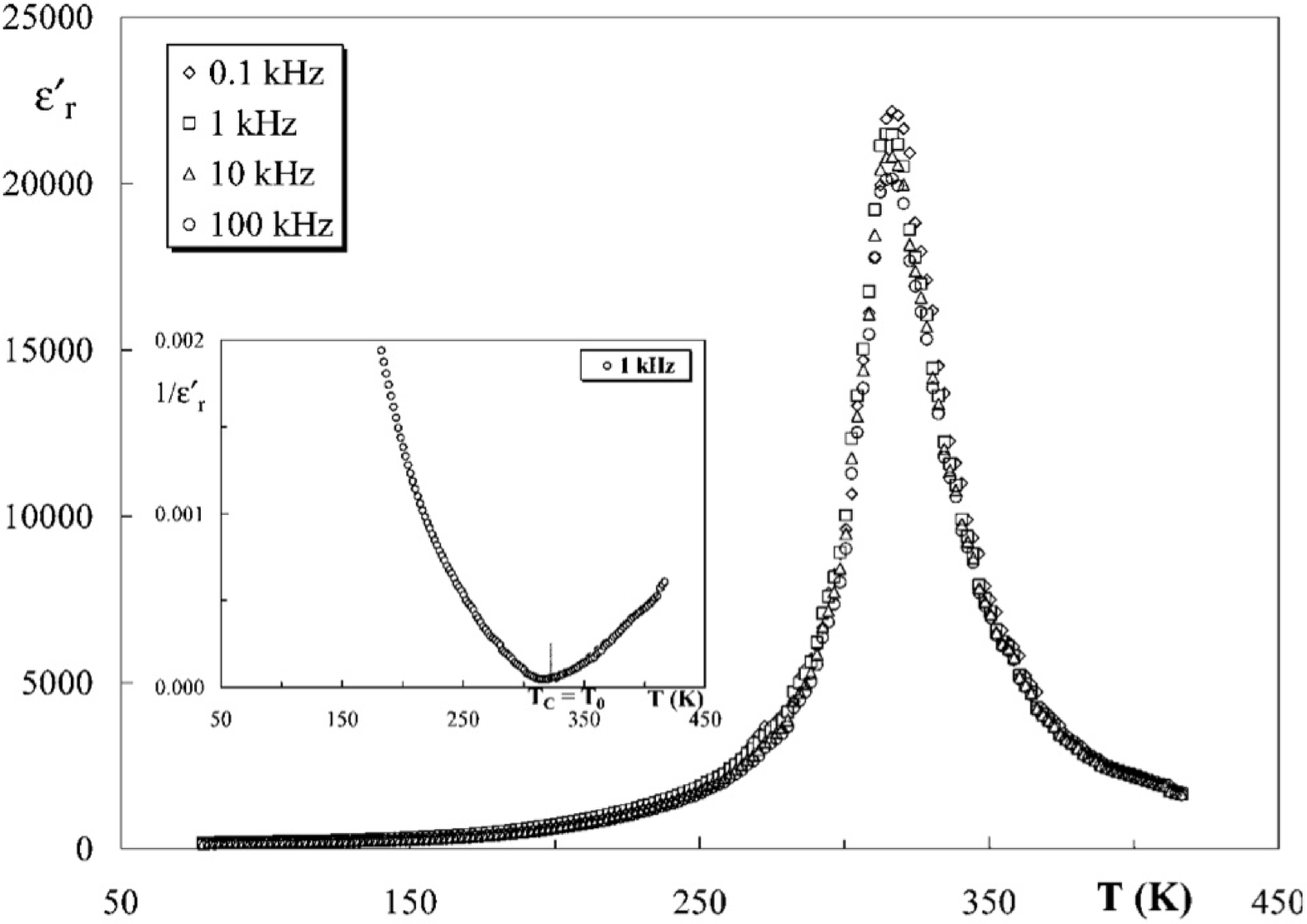}
  \caption{
 AC susceptibilities: (i) heterovalent relaxor ${ \rm {PbMg_{1/3}Nb_{2/3}O_3}   }$  (PMN),  from \cite{Smolenskii_PMN} with permission \copyright Physical Society of Japan (1970);  
 (ii) spin glass {\bf{Cu}}Mn, from \cite{Mydosh-Mulder}, with permission \copyright American Physical Society (1981);
 (iii) homovalent relaxor 
${\rm{BaZr_{0.35}Ti_{0.65}O_3 }}$  (BZT), from \cite{Simon}, with permission \copyright IOP Publishing (2004);
 (iv) ferroelectric ${\rm{BaZr_{0.20}Ti_{0.80}O_3 }}$ from \cite{Simon}, with permission \copyright IOP Publishing (2004)  }
\label{AC_susc}}
\end{figure}
 %
 \begin{figure}
 \includegraphics[width=4.7 in]{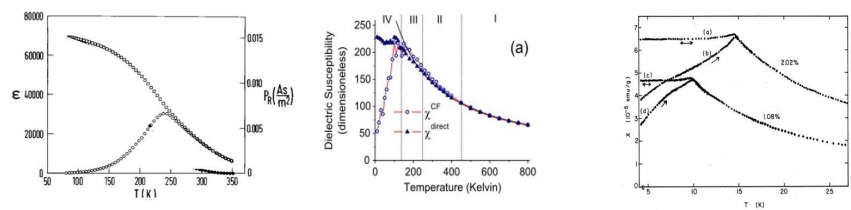}
 \caption{Field cooled (FC) vs. zero-field cooled (ZFC) static susceptibilities; (i) PMN from \cite{Levstik}, with permission \copyright American Physical Society (1998); (ii)${\rm{ BZT}}_{50/50}$ simulation, from \cite{Akbarzadeh2012}, with permission \copyright American Physical Society;(2012) (iii) {\bf{Cu}}Mn spin glass, from \cite{Nagata}, with permission \copyright American Physical Society (1979).
 } 
 \label{FC_ZFC}
 \end{figure}

 However, it is instructive to consider first the more recently discovered alloy ${\rm{Ba {Zr_{x}Ti_{1-x}} O_{3}}}$ (BZT)\cite{Simon}, in which the Ti and Zr  B-ions are homovalent (both 4+) and so without the extra, potentially complicating, charge- (and hence effective field-) disorder of PMN.  The AC susceptibility for BZT with $x=0.35$ is shown in Fig. ({\ref{AC_susc}}(iii). It looks very similar to those shown for PMN and {\bf{Cu}}Mn, suggesting similar physics. (Again, the average crystal structure remains cubic perovskite and without global electric order.) The preparation-dependence of static susceptibilities also look similar, with a separation between field cooled and zero field cooled susceptibilities onsetting at the transition and growing as the temperature is reduced; see Fig ({\ref{FC_ZFC}).

Noting that pure {$\rm{BaZrO}_3$}
is experimentally found not to exhibit ferroelectricity, in contrast to {$\rm{BaTiO}_3$}, one can already deduce that the local harmonic displacement coefficient of ${\rm{Zr^{4+}}}$ 
must be significantly greater than that of ${\rm{Ti^{4+}}}$, too large for the interaction terms to overcome it
\footnote{Note that, since ${\rm{Zr^{4+}}}$ and  ${\rm{Ti^{4+}}}$ are homovalent, the strengths of the ionic Coulombic interactions in BT and BZ are essentially the same.} \footnote{In fact the coefficients can be calculated from numerical first-principles quantum mechanics. Another crude indicator comes from comparison of ionic radii.}.

The relaxor phase of BZT, as discussed above,  is straightforwardly explainable by analogy with the induced-moment discussion above for a spin glass phase in {\bf{Rh}}Co, with Zr the analogue of Rh, Ti of Co, and the relaxor the analogue of a spin glass, with in-cell ionic displacements analogues of local magnetizations; see Figs. (\ref{BZT_schematic}) and (\ref{Maiti})
Simulations of ${\rm{ BZT}}_{50/50}$ using a model as discussed above have been performed by \cite{Akbarzadeh2012} (FC/ZFC; Fig (\ref{FC_ZFC})(ii)) and \cite{Wang2016} (AC susceptibility), in good accord with experiments. 
}
\begin{figure}[ht]
\sidebyside
{\includegraphics[width=2.2in, height=1.3in]{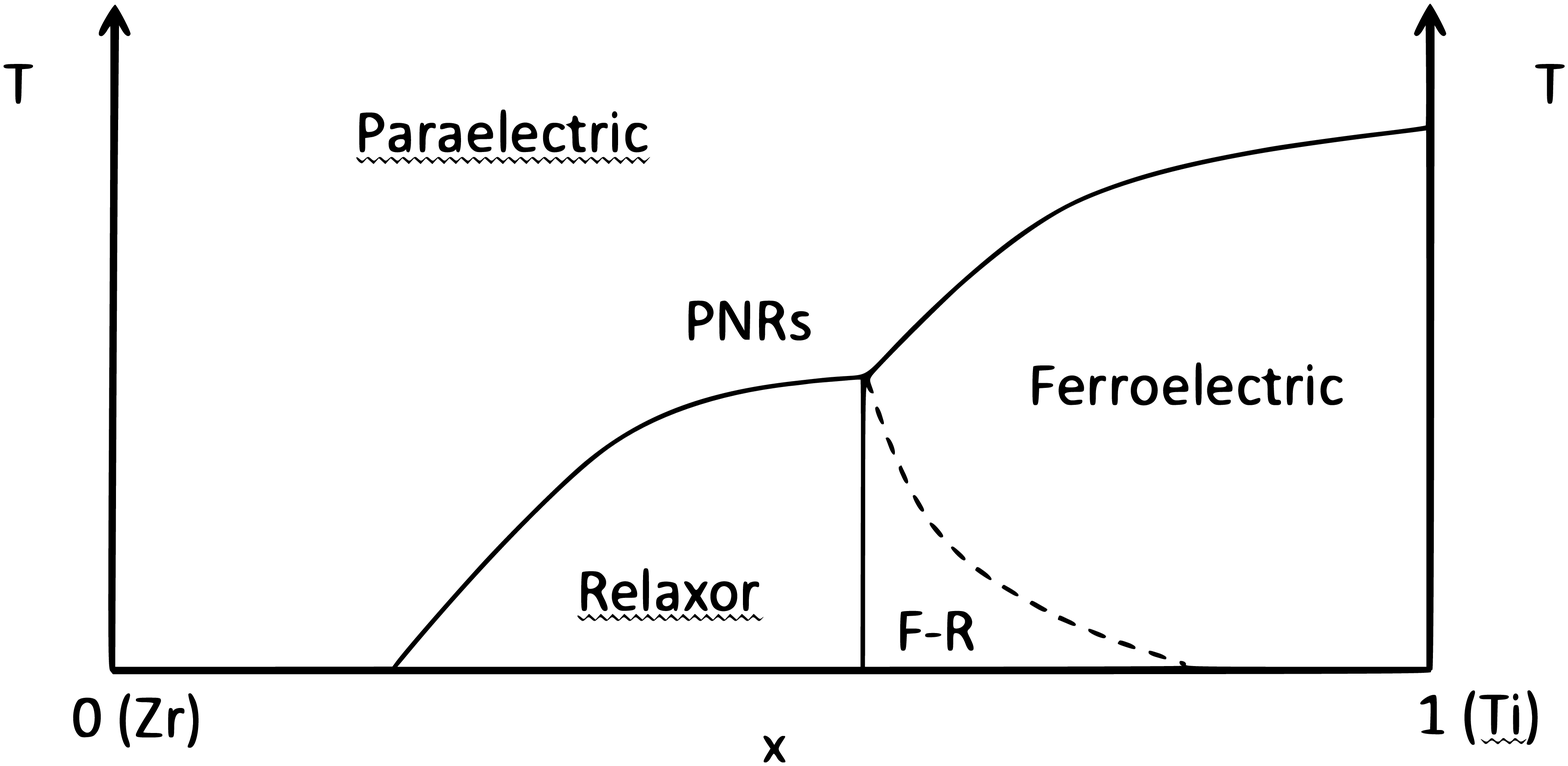}
 \label{BZT_schematic}
 }
 {
\includegraphics[width=2.0in, height=1.4in]{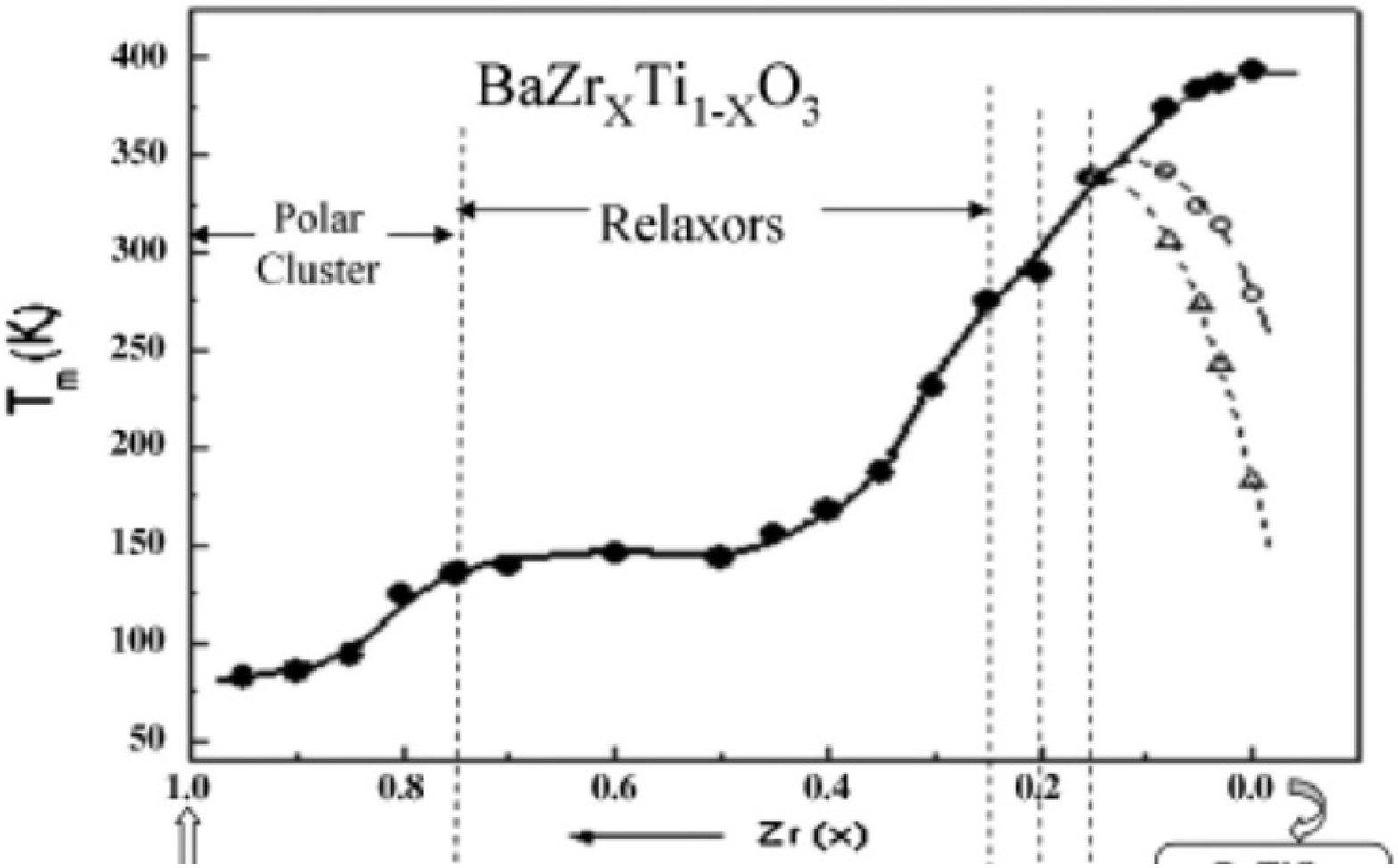}
\label{Maiti}}
\caption{ {\bf{Left}}: Heuristic phase diagram for BZT, by analogy with spin glass. PNRs refer to locally correlated statistical clusters (polar nano-regions) above the extended cooperative transition \cite{Sherrington2014}. F-R refers to a possible mixed ferroelectric-relaxor. \newline  {\bf{Right:}} Experimental phase diagram for BZT,  from  \cite{Maiti}, with permission \copyright American Ceramics Society (2008)}
\end{figure}

It is also interesting to compare  the frequency-dependences of the location of the peaks in experimental AC susceptibility curves of ${\rm{BaZr_{x}Ti_{1-x}O_3 }}$ for  concentrations of the Zr ions above and below the  critical concentration separating  ferroelectric and relaxor.
The ferroelectric  transition peak temperature of the AC susceptibility for  $x=0.20$
(ferroelectric; Fig (\ref{AC_susc})(iv)) is essentially frequency-independent, as  for a normal pure ferroelectric transition, while that for $x=0.035$ (relaxor; Fig({\ref{AC_susc})(iii))
 is highly frequency-dependent, as indeed are the peaks for spin-glass onset in low concentration {\bf{Cu}}Mn (Fig({\ref{AC_susc}(ii)) and heterovalent relaxor PMN (Fig({\ref{AC_susc}(i)). It seems likely that the greater dispersion of the relaxors and spin glass, compared with their ferro counterparts, are consequences of the rugged and chaotically-evolving  {\it{free energy landscape}} that underlies RSB.

Let us now return to   
the  original, canonical and much studied relaxor  PMN. 
It differs from BZT in that its site-occupation  disorder is heterovalent, with the two types of B ions carrying different charges,
Mg being 2+ and Nb 5+, rather than the 4+ of the normal pure ($\rm{ABO_3}$) template.
Being also randomly distributed, they effectively lead to extra quenched random fields in  an extension of eqn.(\ref{S-Mihill}), in three dimensions.

Noting that ${\rm{Mg^{2+}}}$ has a similar ionic radius to ${\rm{Zr^{4+}}}$ and ${\rm{Nb^{5+}}}$ a similar ionic radius to ${\rm{Ti^{4+}}}$, one can expect some B-ordering tendencies in PMN similar to those in BZT  \cite{Sherrington2014b}. However, since ${\rm{Pb{Zr_{x}Ti_{1-x}} O_{3}}}$ (PZT) has not been observed to exhibit relaxor behaviour, the random fields appear to be also necessary for the favouring of spin glass behaviour in PMN\footnote{An experimentally-based case for the important role of random fields in PMN was originally made by \cite{Westphal1992}.} \footnote{We might also note that ${\rm{Pb^{2+}}}$  is significantly more easily displaceable than ${\rm{Ba^{2+}}}$; in ferroelectric BT the relative  displacements of  Ba:Ti is 1:4   while in PT the relative displacements of Pb:Ti is 7:3, so that the undiluted A-site (Pb) ions play the dominant role in PMN, in contrast to the diluted B-site Ti-ions in BZT. This is in accord with the smaller ionic radius of ${\rm{Pb^{2+}}}$ compared with ${\rm{Ba^{2+}}}$}. 
This further raises the issue of possible \mbox{(quasi-)}spin-glass behaviour in even-simpler systems with a combination of  frustrated but non-random interactions and quenched local disorder only through  random fields.
Given that truly 3-dimensionally random local magnetic fields are impossible to produce experimentally, heterovalent relaxor alloys may provide fruitful alternative {\it{experimental laboratories}} for the study of their consequences.

A third feature observed experimentally in relaxors, and  considered as one of their key characteristics, is of clusters of locally ordered regions above the (susceptibility-peak) transition temperature, generally referred to as `polar nano-regions' (PNRs). Within the picture outlined above they are simply explainable for homovalent systems (such as BZT) as quench-statistically-occurring localized states
in an analogue of  Anderson localization \cite{Anderson1958} in which a  `mobility edge' determines the separation of cooperative extended order and insufficiently correlated PNRs, while a density-of-states 'band edge' emulates the limit of PNR observability \cite{Sherrington2014, Sherrington2018}.

A  minimal SK-analogue induced-moment  model is
\begin{equation}
H= \sum_{i}
(r m_{i}^2 + u m_{i}^4)
 -\sum_{(ij)} J_{ij} m_{i} m_{j},
\label{Solvable soft SG}
\end{equation}
\cite{Sherrington2015}, with the \{$J_{ij}$\} drawn independently randomly from the infinite-ranged SK distribution, offering the possibility of exact solution\footnote{It has also recently been proposed and studied in connection with 
low temperature vibrational excitations of structural glasses \cite{Bouchbinder2021}.}. However, since all the eigenstates of 
the SK-distribution for \{$J_{(ij)}$\} are extended, no PNRs should result.

In summary, in this sub-section we have drawn attention to displacive relaxor ferroelectrics that display features that are reminiscent of spin glasses\footnote{For a discussion of pseudo-spin glass-like behaviour in material systems with pre-formed moments, see \cite{Binder-Reger, Sherrington-Osaka, Sherrington2019}.}. Although these features were observed already six decades ago and have led to considerable practical applications, their understanding nevertheless remains controversial. We have argued for an explanation as induced pseudo-spin glasses. Furthermore, by comparison of two different relaxor examples, homovalent and heterovalent, we speculate on an important role for quenched random fields in systems with spatially frustrated interactions, and on heterovalent relaxors as potentially valuable experimental ‘laboratories’ that do not exist for magnetic systems.

\section{Short-range and lattice models}

The discussions above have been principally in terms of range-free systems. Real experimental systems are normally lattice-based in physical dimensions, normally $d=3$, sometimes with relatively short-ranged, but still frustrated,  interactions ({\it{e.g}} {$\rm{Eu_{x}Sr_{1-x}S}$}), sometimes long-ranged but decaying with separation ({\it{e.g}} {\bf{Au}}Fe, {\bf{Cu}}Mn). There has been much theoretical investigation concerning whether such systems, particularly short-ranged, can have RSB, whether there is a lower critical dimension for an AT transition in a field  and, if so, what. These issues will not however be further discussed here, but left to later chapters, as will many applications and extensions of concepts derived from infinite-ranged modelling.

\section{Conclusion}
We have introduced several  extensions of the Ising SK spin glass model, to different kinds of variables/spins/pseudo-spins and several types of interactions, mostly still randomly drawn from range-free intensive distributions, indicating several of the new features that have been exposed by theorists' `blue sky' curiosity, mainly from the first few decades of spin glasses. We have also noted two examples of conceptual and technical transfers, to graph partitioning and relaxor ferroelectrics.

Later chapters will extend the developments and expose much further theoretical enlightenment  and fruitful conceptual and practical progress in application to problems far beyond understanding the metallic alloys that stimulated the theory of spin glasses half a  century ago.

\bibliography{DSJA}

\bibliographystyle{ws-book-van} 

\printindex

\end{document}